\documentclass{aa}   
\usepackage{graphicx} 
\usepackage{txfonts} 
\usepackage[]{natbib} 
\begin{document} 
   \title{A wide-field \ion{H}{i} mosaic of Messier 31} 

   \subtitle{II. The disk warp, rotation and the dark matter halo} 

   \author{Edvige Corbelli 
          \inst{1} 
          \and 
          Silvio Lorenzoni 
          \inst{1} 
          \and 
          Rene Walterbos 
          \inst{2}          
          \and 
          Robert Braun 
          \inst{3} 
          \and 
          David Thilker  
          \inst{4} 
          } 


   \institute{ 
              INAF-Osservatorio Astrofisico di Arcetri, Largo E. Fermi, 5 -   
              50125 Firenze - Italy\\ 
              \email{[edvige, silvio]@arcetri.astro.it} 
         \and 
              Department of Astronomy, New Mexico State University, 
              P.O. Box 30001, MSC 4500, Las Cruces, NM 88003, USA \\ 
              \email{rwalterb@nmsu.edu} 
         \and  
              CSIRO-ATNF, PO Box 76, Epping, NSW 2121, Australia\\ 
              \email{Robert.Braun@csiro.au} 
         \and 
              Center for Astrophysical Sciences, Johns Hopkins 
              University, 3400 North Charles Street, Baltimore, MD 
              21218, USA \\ 
              \email{dthilker@pha.jhu.edu} 
             }

   \date{Received; accepted} 


  \abstract 
   {} 
{We test cosmological models of structure 
formation using the rotation curve of the nearest spiral galaxy, M31, 
determined using a recent deep, full-disk 21-cm imaging survey smoothed to 
466~pc resolution.} 
{We fit a tilted ring model to the HI data from 8 
to 37~kpc and establish conclusively 
the presence of a dark halo and its density distribution via dynamical analysis 
of the rotation curve.} 
{The disk of M31 warps from 25~kpc 
outwards and becomes more inclined with respect to our line of
sight. Newtonian dynamics without a dark matter halo provide a very poor 
fit to the rotation curve.  
In the framework of modified Newtonian dynamic (MOND) however the 21-cm  
rotation  curve is well fitted by the
gravitational potential traced by the baryonic matter density alone.
The inclusion of a dark matter halo with a density 
profile as predicted by hierarchical clustering and structure 
formation in a $\Lambda$CDM cosmology makes the mass model in newtonian
dynamic compatible with the rotation curve data. The dark halo concentration 
parameter for the best fit is $C=12$ and its total mass is 1.2~10$^{12}$~M$_\odot$.
If a dark halo model with a constant-density core is considered, the
core radius has to be  larger  than 20~kpc in order  
for the model to provide a good fit to the data.  We
extrapolate the best-fit $\Lambda$CDM and constant-density core mass models
to very large galactocentric radii, comparable to the size of the dark matter halo. 
A comparison of the predicted mass with the M31 mass determined at such large radii
using  other dynamical tracers, confirms the validity of our results. In particular
the $\Lambda CDM$ dark halo model which best fits the 21-cm data
well reproduces the mass of M31 traced out to 560~kpc. Our best estimate 
for the total mass of M31 is  
1.3$\times 10^{12}$~M$_\odot$, with 12$\%$ baryonic fraction and only
6$\%$ of the baryons in the neutral gas phase.} 
{}


   \maketitle%

\section{Introduction} 

Rotation curves of spiral galaxies are fundamental tools to study the
visible mass distributions in galaxies and to infer the properties of
any associated dark matter halos. These can then be used to constrain
cosmological models of galaxy formation and evolution. Great effort
has been devoted in recent years to test theoretical predictions of
cosmological models regarding the detailed structure of dark matter halos via
observations on galactic and sub-galactic scales.  Knowledge of the
halo density profile from the center to the outskirts of galaxies is
essential for solving crucial issues at the heart of galaxy formation
theories, including the nature of the dark matter itself.  Numerical
simulations of structure formation in the flat Cold Dark Matter
cosmological scenario (hereafter $\Lambda$CDM) predict a well defined
radial density profile for the collisionless particles in virialized
structures, the NFW profile \citep{1996ApJ...462..563N}. 
While there
is a general consensus that hierarchical assembly of $\Lambda$ CDM
halos yields ``universal'' mass profiles on a large scale
(i.e. independent of mass and cosmology aside from a simple two
parameter scaling), there is still some controversy on the central
density profile, and on the relative scaling parameters.  For example
\citet{2004MNRAS.349.1039N} proposed the ``Einasto profile'', a
three-parameter formulation, to improve the accuracy of the fits to cuspy
inner density profiles of simulated halos.  The two parameters of the
``universal'' NFW density profile are the halo overdensity and the
scale radius, or (in a more useful parameterization) the halo
concentration and its virial mass. For hierarchical structure
formation, small galaxies should show the highest halo concentrations
and massive galaxies the lowest ones \citep{2007MNRAS.378...55M}. Do
the observations confirm these predictions?  Dwarf galaxies with
extended rotation curves have often contradicted this theory since
central regions show shallow density cores, i.e.  very low dark matter
concentrations
\citep{2004ApJ...617.1059R,2005ApJ...634L.145G,2007MNRAS.375..199G}.
This has given new insights on the nature of dark matter and lead to
discussion on how the halo structure might have been altered by the
galaxy formation process \citep[e.g. ][]{2002MNRAS.333..299G}. Recent
hydrodynamical simulations in $\Lambda CDM$ framework have shown that
strong outflows in dwarf galaxies inhibit the formation of bulges and
decrease the dark matter density, thus reconciling dwarf galaxies with
$\Lambda CDM$ theoretical predictions \citep{2009arXiv0911.2237G}.
Shallow density cores have often been supported also by high  
resolution analysis of rotation curves of spirals and low surface
brightness galaxies \citep{2001ApJ...552L..23D,2004MNRAS.351..903G}.  
However, uncertainties related to the presence of non-circular motion (often
related to the presence of small bars and bulges), observational
uncertainties on the observed velocities, and the possibility of dark matter
compression during the baryonic collapse leave still open the question
on how dark matter is effectively distributed in today's galaxies 
\citep{2000AJ....119.1579V,2003MNRAS.342..199C,2007ApJ...669..315C}.

Bright galaxies, because of the large fraction of visible to dark
matter, do not offer the possibility to trace dark matter very
accurately in the center. Uncertainties related to distance estimates,
to the disk-bulge light decompositions, and the typically limited
extention of the gaseous disks beyond the bright star forming disks
further limit the ability to derive accurate dynamical mass models.
These difficulties can be alleviated in the case of the spiral galaxy
M31 (Andromeda).  Owing to its size, proximity, well known distance,
and to constraints on its structural parameters from the long history
of observations at all wavelengths, Andromeda (the nearest giant
spiral galaxy) offers a unique opportunity to analyze in detail
the mass distribution and the dark halo properties in bright disk
galaxies.  Massive galaxies like M31 can probe dark matter on mass
scales much larger than that of the dwarfs, of order
10$^{12}$~M$_\odot$.

The Milky Way and Andromeda are the most massive 
members of the Local Group. Any estimate of their total mass and of 
the structure of 
their dark matter halos is a requirement for any study of the 
dynamics of the Local Group, its formation, evolution, and ultimate fate  
of its members \citep[e.g.][]{2008MNRAS.384.1459L}.  Difficulties in the 
determination of the Milky Way's mass components, related to the fact 
that our solar system is deeply embedded in its disk, can be 
overcome in the case of M31. 
M31 is known to have a complex merging history. Its multiple nucleus 
\citep{1993AJ....106.1436L} and the extended stellar stream 
and halo \citep[e.g. ][]{2001INGN....5....3I,2008MNRAS.390.1437C} are 
clear signs of a tumultuous life. According to the hierarchical models 
of galaxy formation it is conceivable that M31 has grown by accretion 
of numerous small galaxies. It is likely the most massive member of 
the Local Group \citep[e.g. ][] {2002ApJ...573..597K}. It is therefore 
of great interest to test the other predictions of hierarchical models 
such as the presence and structure of a dark matter halo around it. 
Contrary to dwarf 
galaxies, luminous high surface brightness galaxies such as M31, 
cannot be used to test the central dark matter distribution, not only 
because of the large surface density of baryons which makes it 
difficult to constrain dark matter, but also because of possible 
adiabatic contraction \citep{2002ApJ...573..597K,2008MNRAS.389.1911S}. 
An extended and well defined rotation curve can instead be complemented 
by the extensive information 
now available on the M31 stellar disk, stellar stream, globular clusters 
and orbits of Andromeda's small satellite galaxies to establish
the dark matter density profile at  large galactocentric distances.
And this is one of our goals.

M31 was one of the first galaxies where Slipher (in 1914) found 
evidence of rotation and also the first galaxy to have a published 
velocity field \citep[][ and references therein]{2001ARA&A..39..137S}.
Using the M31 rotation curve, \citet{1939LicOB..19...41B}
was the first person to advocate unseen mass at large radii in a galaxy.
Since then much effort has been devoted to study the rotation curve 
of M31 and to understand the relation between the light and the mass 
distribution. Despite a century of dedicated work, there are 
still many unsettled questions concerning the shape of the M31  
rotation curve, the contribution of visible and dark matter to it, 
and the changing orientation of the M31 disk. 
Detailed HI surveys with single dish or synthesis observations 
\citep[e.g. ][]{1977MNRAS.181..573N,1980A&AS...40..215C, 
1982A&AS...49..745B,1983MNRAS.205..773U,1984A&A...141..195B,
1991ApJ...372...54B} have been  
analyzed to find local kinematical signatures of spiral arm 
segments, of the interaction with M32 or of a warped disk or ring. 
Even though many authors have pointed out  the presence of a warp  
in M31, i.e. of a systematic deviation of the matter distribution from 
equatorial symmetry, a complete quantitative analysis of the parameters  
of such a distortion at large galactocentric radii is still missing. 
Previous modelling of the HI warp has been based on a combination of 
high resolution inner disk HI data and much lower resolution and 
sensitivity outer disk data. The models assumed a rotation curve but  
no independent fit of rotation curve and warping of the disk has been  
attempted \citep[e.g. ][]{1979A&A....75..311H,1984A&A...141..195B}. 
Some papers \citep[e.g. ][]{2006ApJ...641L.109C} analyze 
the extended rotation curve of M31 using only HI data  
along the direction of the optical major axis, without considering the 
possibility of a warped disk. 
Only very recently \citet{2009ApJ...705.1395C} use deep 21-cm survey 
of the M31 based on high resolution synthesis observations to model the
warp and the rotation curve simultaneously.

Our first aim is to use the new WRST HI survey of M31 
\citep{2009ApJ...695..937B} to define the 
amplitude and orientation of the warp using a tilted ring model.  A 
set of free rings will be considered for which the following 
parameters need to be determined: the orbital center, the systemic 
velocity, the inclination and position angle with respect to our line of 
sight, and the rotational velocity. The geometric properties of the 
best fitting tilted ring model will then be used to derive the 
rotation curve from the 21-cm line observed velocities.  The final 
goal will be to use the rotation curve for constraining the baryonic 
content of the M31 disk and the presence and distribution of dark 
matter in its halo through the dynamical analysis. 

Our recent deep wide-field HI imaging survey  reaches a maximum 
resolution of about 50~pc and 2~km~s$^{-1}$ across a 
95$\times$48~kpc$^2$ region. This makes our database the most detailed 
ever made of the neutral medium of any complete galaxy disk, including 
our own. Observations and data reduction are described in 
\citet{2009ApJ...695..937B} (hereafter Paper I). In Paper I we 
analyzed HI self-absorption features and find opaque atomic gas 
organized into filamentary complexes. While the gas is not the 
dominant baryonic component in M31, we take these opacity 
corrections into account in determining the dynamical contributions of 
the various mass components to the M31 rotation curve.  In this paper 
we use the data at a resolution of 2~arcmin (457~kpc) in order to gain sensitivity  
in the outermost regions. At this spatial resolution, we reach a brightness 
sensitivity of 0.25~K. Considering a typical signal width of 
20~km~s$^{-1}$ our sensitivity should be appropriate for detecting HI 
gas at column densities as low as 10$^{19}$~cm$^{-2}$. 

In Section 2, we describe the modeling 
procedures for determining the disk warp in M31 and discuss the resulting  
disk orientation. In Section 3, we determine the rotation curve  
and the uncertainties associated with it. Various dynamical mass models for 
the rotation curve fit are introduced and discussed in Section 4. 
We determine the total baryonic and dark mass of this galaxy.  
Together with complementary data at very large galactocentric radii, we 
confirm the predictions of $\Lambda$CDM cosmological models.
Section 5 summarizes the main results of this paper. 

We assume a distance to M31 of 785~kpc throughout, as derived by  
\citet{2005MNRAS.356..979M} (D=$785\pm 25$~kpc).


\section{Tilted rings: modeling procedures} 

For a dynamical mass model of a disk galaxy it is necessary to 
reconstruct the tri-dimensional velocity field from the velocities 
observed along the line of sight. If velocities are circular and 
confined to a disk one needs to establish the disk orientation for 
deriving the rotation curve, i.e. the position angle of the major axis 
(PA), and the inclination of the disk with respect to the line of 
sight ($i$).  If the disk exhibits a warp these parameters vary with 
galactocentric radius. This is often the case for gaseous disks 
which extend outside the optical radius and which often show a 
different orientation than the inner one.  Our attempt to 
understand the kinematics of M31 is done performing a tilted ring 
model fit to the data, under the assumption of circular motion. 
Because of this assumption we will use the tilted ring model outside 
the inner 8~kpc region, i.e. where deviations of gas motion from circular 
orbits are expected to be small.  We will not consider local 
perturbations to the circular velocity field such as those due to 
spiral arms. The comparison between the velocities predicted by a 
tilted ring model and the data is done over all azimuthal angles and 
not only around the major axis. This will average out spiral arm 
perturbations. 

\begin{figure} 
\includegraphics[width=\columnwidth]{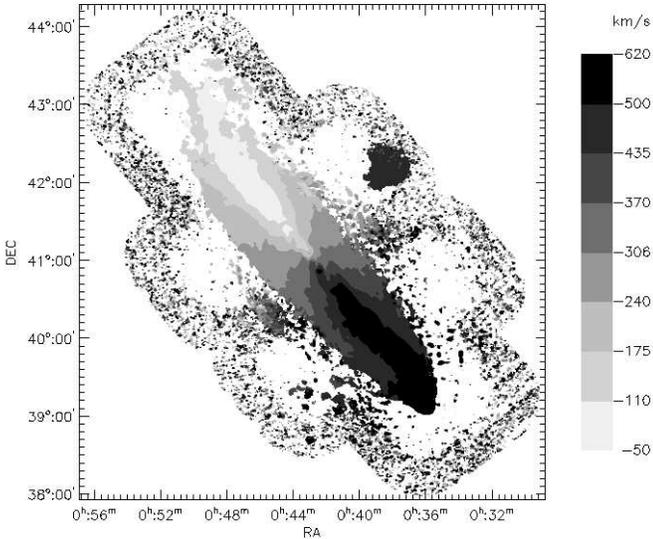} 
\caption{The first moment map. The intensity-weighted mean velocity has been 
computed  
from the 120~arcsec data cube at a spectral resolution of 2~km~s$^{-1}$,  
using a 4-$\sigma$ blanking in the data cube at 18.5 km/s resolution
to determine what is included in the integral.} 
\label{mom1} 
\end{figure} 

In Figure~\ref{mom1} we show the first moment map (i.e. the intensity-weighted
mean velocity along the line of sight) of our 21-cm Andromeda survey
\citep[see ][ for more details]{2009ApJ...695..937B}. 
Because of the large angular extent of Andromeda, it is necessary to 
consider the correct transformation between angular and cartesian 
coordinates in order to derive galactocentric distances.  A detailed 
description of the spherical trigonometry involved can be found in the 
literature \citep[e.g. ][]{2001AJ....122.1807V}.  We shall summarize 
here the most important formulae. Note that these take into account 
the variations in the distance between us and the Andromeda regions 
that do not lie along the line of nodes (the disk on the near side 
(West) half is closer to us than on the far side (East) half). 
Consider a point at a given right ascension and declination 
($\alpha,\delta$) in the Andromeda disk having inclination $i$ and 
position angle $P.A.=\theta$. The distance between the observer and 
this point is $D$. The center of the galaxy has right ascension, 
declination ($\alpha_0,\delta_0$) and distance $D_0$ from the 
observer.  Consider $\phi$ as the angle between the tangent to the 
great circle on the celestial sphere through ($\alpha,\delta$) and 
($\alpha_0,\delta_0$) and the circle of constant declination 
$\delta_0$ measured counterclockwise starting from the axis that runs 
in the direction of decreasing R.A. at constant declination $\delta_0$ 
(in practice $\phi=P.A.+90^\circ$).  The value of $\phi$ along the 
line of nodes (for points along the major axis) is $\Theta$ 
($\Theta=\theta+90^\circ$).  The angular distance between 
($\alpha,\delta$) and ($\alpha_0,\delta_0$) is the second angular 
coordinate called $\rho$.  We shall work in a Cartesian coordinate 
system ($x,y,z$) which has its origin at the galaxy center, the 
$x$-axis antiparallel to R.A., the $y$-axis parallel to declination 
axis and the $z$-axis toward the observer.  The coordinates of the 
observed pixels in this system will be: 

\begin{equation} 
x =D\ \hbox{sin}\rho\ \hbox{cos}\phi \qquad 
y =D\ \hbox{sin}\rho\ \hbox{sin}\phi  
\end{equation} 

\noindent 
where 

\begin{equation} 
D =D_0\ \hbox{cos} i \ /\lbrack \hbox{cos} i\ \hbox{cos}   
\rho\ - \hbox{sin} i\ \hbox{sin} \rho\ \hbox{sin} (\phi-\Theta)  \rbrack  
\end{equation} 

\noindent 
Because of the warp, $i$ and $\theta$ are not constant. This implies 
that for a given ring model the numerical code starts with a guessed 
matrix for $i$ and a guessed matrix for $\theta$ over the observed pixels.  
Once galactocentric distances are determined it checks the matrices with  
the tilted ring model and iterates.  Positions and velocities 
of a ring at a given inclination and position angle ($i_i,\theta_i$) 
can be easily derived into this reference system  by a two angle 
rotation ($i_i,\theta_i$) of the Cartesian coordinate system which has 
the ring in the ($x',y'$) plane \citep[e.g. ][]{1989A&A...223...47B}. 

The tilted ring model is (as usual) based on a number of geometrical and 
kinematical parameters relative to the HI disk.  To represent the HI 
distribution in M31 we consider 110 rings between 10 and 35~kpc in radius. 
Each ring is characterized by its radius $R$ and by 6 
additional parameters: the circular velocity $V_c$, the inclination 
$i$, the position angle $\theta$, the systemic velocity $V_{sys}$ and 
the position shifts of the orbital center with respect to the galaxy center 
($\Delta x_c, \Delta y_c$). These last 3 parameters allow the rings to 
be non concentric and the gas at each radius to have a 
velocity component perpendicular to the disk (such as that given by 
gas outflowing or infalling into the disk). If the average value of this
velocity component is uniform and non zero over the whole disk it implies 
that effectively the systemic velocity determined from the gas 
velocity field is different from the assumed one. We shall also consider 
the case of a tilted 
ring model with uniform values of $V_{sys},\Delta x_c, \Delta y_c$. 
To test the model each ring is subdivided into segments 
of equal area smaller than the spatial resolution of the dataset in 
use. We convolve the emission from various pieces
with the beam  pattern at each location to compute the velocity along 
the line of sight 
and its coordinates in a rest frame, defined above. In the same rest 
frame we consider the 21-cm dataset. The telescope beam shape for the 
dataset we use  (a smoothed version of the original data) 
can be well described by a Gaussian function with full width half 
maximum (FWHM) of 2~arcmin. With this method we naturally account for 
the possibility that the intensity-weighted mean velocity in a pixel 
might result from 
the superposition along the line of sight of emission from different 
rings.  For each position in the map we compare the observed 
velocity $V^{obs}$ with the velocity predicted by the model $V^{mod}$. 
The observed velocity is the mean 
velocity of the HI gas at each position.  The assignment of a 
measure of the goodness of the fit in this modeling procedure is done 
using a $\chi^2$ test on the difference between $V^{obs}$ and 
$V^{mod}$. 
The noise is uniform in our map 
\citep[see ][]{2009ApJ...695..937B} and we assign uniform uncertainties 
$\sigma_i$ to the observed values $V^{obs}$. In order to keep the model 
sensitive to variations of parameters
of the outermost rings, each pixel in the map is assigned equal 
weight. Pixels with higher or lower 21-cm surface brightness 
contribute equally to determine the global goodness of the model fit. 
Since the velocity channel 
width  of our survey is about 2~km~s$^{-1}$, we arbitrarily set
$\sigma_i$ equal to the width of 3 channels (6~km~s$^{-1}$). 
This is simply a scaling factor.
The equation below defines the reduced $\chi^2$ which we will use
throughout this paper

\begin{equation} 
\chi^2={1\over N - \nu} \sum^N_{i=1}(V^{mod}_i-V^{obs}_i)^2/\sigma_i^2  
\end{equation}  

\noindent 
In our definition of $\chi^2$, 
N is the number of positions with HI column density along the 
line of sight greater than $5\times 10^{19}$~cm$^{-2}$ ($N=31391$) and  
$\nu$ is the number of degrees of freedom.  It is unreasonable to 
consider 110 free rings i.e. each ring with 6 degrees of freedom, not 
only because of the large amount of computer time required to find a 
solution but also because solutions with so many degrees of freedom 
are not very robust (parameter variations for 1 out of 110 rings does
not give sensible variations to the $\chi^2$). On the other hand the use 
of functional forms 
that describe the 6 free parameters as a function of R, each with 2-3 
free parameters, often do not give satisfactory results 
\citep[e.g. ][]{1997ApJ...479..244C}.  Following 
\citet{1997ApJ...479..244C} we constructed a model in which the 
properties of 11 equally spaced rings could be varied 
independently. We shall call these the `free rings'.
The first free ring is centered at 10~kpc, the 
last one at 35~kpc. This is because placing the first/last free ring at
smaller/larger galactocentric distance make them not very stable.
We extrapolate the geometrical parameters of the first/last free
ring for 2~kpc inward/outward when using the tilted ring model
to derive the rotation curve at radii between 8 and 37~kpc. 
The parameters of the 10 rings between 2 free rings 
are found by linear interpolation.  Our procedure is to search  
for a minimum $\chi^2$ value considering the parameters of the
110 rings. 

Given the numerous free parameters ($\nu=66$) it is still unreasonable 
to search for the minimal solution scanning a multidimensional grid 
(of 66 dimensions). We use two procedures to determine the best 
fitting model. In the first procedure (hereafter P1) we apply the 
technique of partial minima. We evaluate $\chi^2$ by varying each 
parameter separately and interpolating to estimate the value for a 
minimum $\chi^2$ for each parameter. We then repeat the procedure over 
a smaller parameter interval around the new solution.  We start the 
minimization with the following initial set of free parameters: 
$i=77^\circ$, $\theta=38^\circ$, $V_{sys}=-300$~km~s$^{-1}$, $\Delta 
x_c=\Delta y_c= 0$ for all free rings.  As an initial guess of $V_{c}$ we 
give the average rotational velocity derived in each free ring by 
deprojecting the observed velocities within a 20$^\circ$ cone of the 
optical line of nodes, $\theta=38^\circ$ (using $i=77^\circ$ and 
$\theta=38^\circ$ for the deprojection).  We carried out several 
optimization attempts under a variety of initial conditions to avoid 
that our partial minima technique carries the solution towards a 
relative minimum. 

\begin{figure} 
\includegraphics[width=\columnwidth]{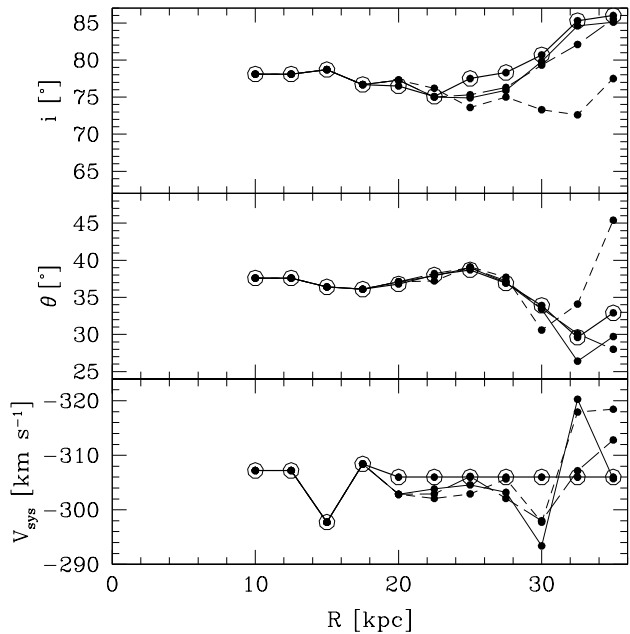} 
\caption{Assortment of ``free ring'' models fit to the HI data according  
to modeling procedure P1. We display 33 of the 66 free parameters: 
the inclination $i$, position angle $\theta$ 
and systemic velocity $V_{sys}$ of the 11 free rings. The 
heavier continuous lines show the best fitting model which gives the minimum 
$\chi^2$ solution (circled points).   
For this case we vary one parameter of a given ring keeping all others at 
the best fitting values. 
Dashed lines indicate some other combinations of free parameters which give 
a $\chi^2$ within within 20$\%$ of the minimum value.} 
\label{ring1} 
\end{figure}

The best fitting ring model is shown in Figure~\ref{ring1} by the 
heavy continuous line. It gives a $\chi^2=2.18$. We display only 3 of  
the 6 free parameters for each free ring, namely the inclination $i$, the 
position angle $\theta$ and the systemic velocity $V_{sys}$. The 
shifts of the orbital centers are small and shown in 
Figure~\ref{shifts}. The minimum $\chi^2$ is obtained for an 
inclination which radially decreases by a few degrees between 10 and 
25~kpc in radius and then increases out to 85~$^\circ$ for the 
outermost ring. The position angle shows marginal variations out to 
25~kpc, then it decreases by about 10~$^\circ$ in the outskirts. 
However, while from 10 to 28~kpc low $\chi^2$ values are obtained for 
combinations of $i$ and $\theta$ not very different than the best 
fitting model, beyond 28~kpc we can find tilted ring models which give 
acceptable fits (with a $\chi^2$ value within 20$\%$ of the minimum) 
with different combinations of $i$ and $\theta$.  The short dashed 
line in Figure~\ref{ring1} shows one model for which the inclination 
of the outermost rings does not increase and their position angle instead varies 
between 30$^\circ$ and 45$^\circ$.  Clearly the outermost rings are not 
very stable, due to a lack of 21-cm emission at these large radii. 
Hence, some care is required when the model is extrapolated to even 
larger radii.   
The single constant value of systemic velocity which minimizes the 
$\chi^2$ is -306~km~s$^{-1}$.  
The errorbars displayed for the best fitting model 
indicate parameter variations for which the $\chi^2$ of the model
is within 20$\%$ of the minimum. This is done varying 
just one parameter at a time with respect to the combination of parameters 
which gives the minimum $\chi^2$.  The rotational velocities $V_c$  
obtained from the fit will 
be discussed in the next Section where we compare them to the  
rotation curve of Andromeda derived from the first moment map
using the geometrical 
parameters and $V_{sys}$ of the best tilted ring model.

\begin{figure} 
\includegraphics[width=\columnwidth]{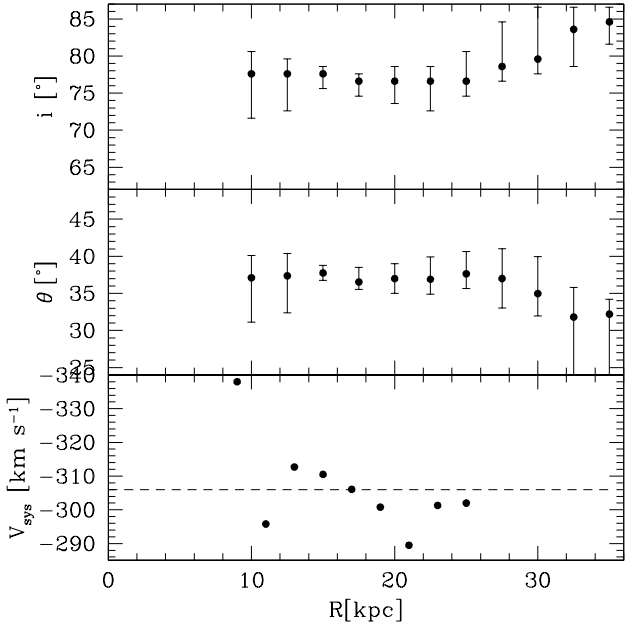} 
\caption{Best fitting P2 solution for $i$ and $\theta$ obtained when 
$V_{sys}=-306$~km~s$^{-1}$, $\Delta x_c =0$, $\Delta y_c =0$. All 
combinations of the 12 free parameters for the outermost 4 free rings 
have been considered. For the innermost 7 free rings instead the minimum 
values have been obtained by varying parameters of one ring at a 
time. Errorbars indicate parameter variations for which  
the $\chi^2$ value of the model is within 20$\%$ of the minimum. The 
lower panel shows the values of $V_{sys}$ determined by the first 
order moment map (see text for details) between 9 and 25~kpc, whose 
average value is -306~km~s$^{-1}$.} 
\label{ring2} 
\end{figure}

In  a second procedure (hereafter P2) we searched for a minimum by 
neglecting the radial variations of $V_{sys},x_c,y_c$. We set $\Delta 
x_c=\Delta y_c=0$ and $V_{sys}=-306$~km~s$^{-1}$.  We have determined 
this value of $V_{sys}$ as described below.  In M31, the integrated 21-cm  
profile shows that the North-East receding side has more gas than the 
South-West approaching side (their ratio being 1.13), at intermediate 
velocities between the central one and the velocity of maximum emission.  
This is 
likely due to the disturbance caused by interaction with M32. 
Hence we cannot simply compute the systemic velocity by averaging the 
observed mean velocities and weighting these with the 21-cm surface 
brightness. Considering the galaxy disk between 10 and 25~kpc with an 
average inclination of 77.7$^\circ$ and position angle 38$^\circ$ we 
compute the average systemic velocity in rings. We weight the data 
with the surface brightness intensity. For points south of the minor axis we 
multiply their weight by the ratio of the HI mass in the northern side 
to that of the southern side in that the ring. The resulting V$_{sys}$ is 
shown in Figure~\ref{ring2}. Averaging the  systemic velocities over all 
rings we thus obtain a value of -306~km~s$^{-1}$. We searched for a minimal 
solution scanning a multidimensional grid (12 dimensions) 
corresponding to parameter variations of the outermost 4 free 
rings. The initial grid of variations for the inclination and position 
angle considered are $\pm 15^\circ, \pm 10^\circ, \pm 5^\circ$ around 
the standard values of the inner disk ($i=77.6^\circ$, 
$\theta=38^\circ$).  The maximum velocity shifts considered with respect to 
the initial values of $V_c$ are $\pm 25,15,10,5$~km~s$^{-1}$.  
We then consider finer grids for the outermost 4 rings. We have 
compared the observed velocities to the velocities predicted by tilted 
ring models using all possible combinations of parameters for the 
outermost 4 free rings.  For the 8 innermost 
free rings, parameter variations have been consider only in a 
3-dimensional space, namely the minimum $\chi^2$ has been found for 
one ring at a time, considering all combinations of $i,\theta,V_c$ for 
that particular ring.  As shown in Figure~\ref{ring2} the 
resulting free parameters for the best fitting tilted ring model are 
very similar to those obtained using the first method. 

We define the residual velocities as

\begin{equation} 
res=V_{mod}-V_{obs} 
\end{equation} 

where $V_{mod}$ are the circular velocities $V_c$, as parameterized in 
the tilted ring model, projected along the line of sight and convolved 
with the beam function. The observed velocities $V_{obs}$ are the mean 
velocities detected along the line of sight. 

Residual velocities are generally smaller than 10~km~s$^{-1}$ and no
characteristic patterns are visible across the galaxy.  
It is worth noticing that, if we trace the major axis position angle by 
a close inspection of the first moment map, the position angle seems 
about constant at $\sim$ 38$^\circ$ out to galactocentric distances 
of 28~kpc, then it decreases to about 32$^\circ$ at 32~kpc and finally 
it increases again back to 38$^\circ$ at 38~kpc.  Given the 
consistency between the best fitting model for the first and second 
procedure (Figure~\ref{ring1} and Figure~\ref{ring2}) and also between these 
tilted ring models and the inspection of the database, we will not 
consider models in which the position angle increases much above 
38$^\circ$ for the outermost rings in the rest of this paper. 

\subsection{The NEMO results} 

\begin{figure} 
\includegraphics[width=\columnwidth]{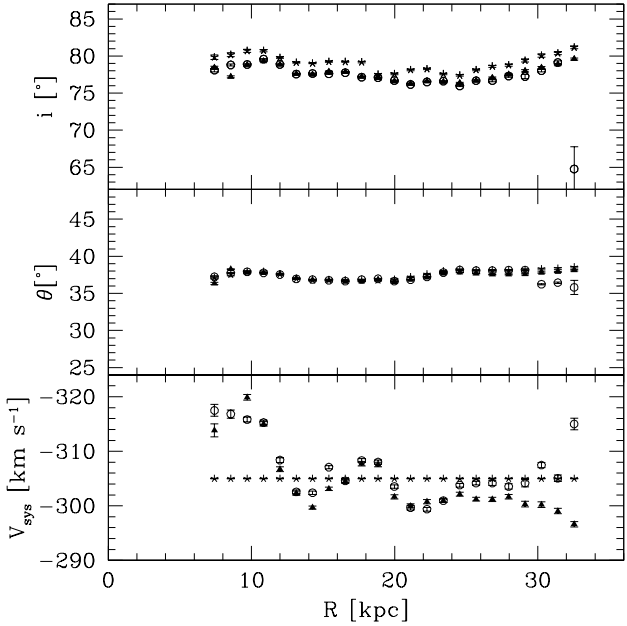} 
\caption{Best fitting solutions obtained with the task ROTCUR in the 
NEMO package (P3). The resulting position shifts of orbital centers 
are very small and hence are not shown.  The open circles indicate the 
best fitting parameters when we exclude data within a 20$^\circ$ angle 
around minor axis in the plane of the galaxy and the data is weighted 
into the least squares solution with the cosine of the galactic angle 
away from the major axis. The triangles indicates the solution 
computed for all data (including points along minor axis) and for 
uniform weight. The star symbols show the best fitting solution 
obtained by keeping the orbital centers fixed, and the systemic velocity 
equal to -305~km~s$^{-1}$, the average value obtained when we allow it 
to vary radially.} 
\label{ring3} 
\end{figure}

\begin{figure} 
\includegraphics[width=\columnwidth]{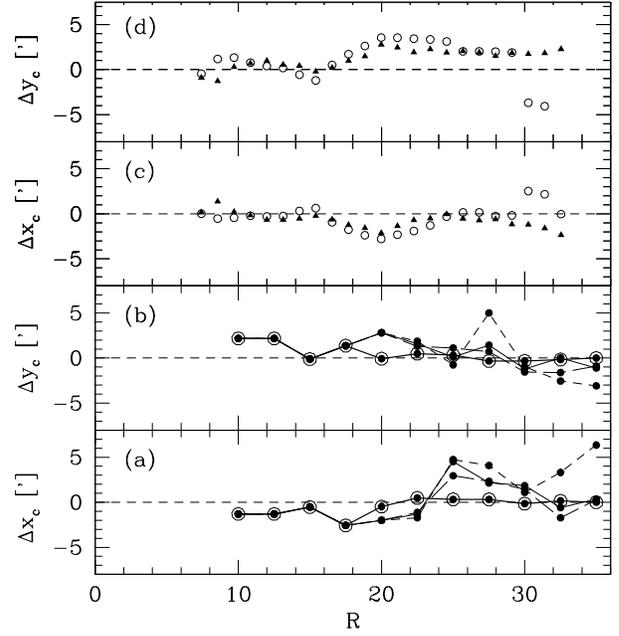} 
\caption{Orbital center shifts in arcmin as derived from tilted ring 
model fits.  Positive values of $\Delta x_c$ correspond to orbital 
centers shifted towards decreasing R.A. values with respect to 
10.6846853$^\circ$ which is Andromeda's center.  Positive values of 
$\Delta y_c$ correspond to orbital centers shifted towards increasing 
Dec. values with respect to the central value of 41.2690374$^\circ$. In panels 
$(a)$ and $(b)$ we show the values corresponding to tilted ring models 
P1 in Figure~\ref{ring1} where we have considered these 
variables. Similarly in panels $(c)$ and $(d)$ the values derived for 
models P3 shown in Figure~\ref{ring3}.} 
\label{shifts} 
\end{figure}

To check our results for the best fitting tilted ring model we also 
used the standard least-square fitting technique developed by 
\citet{1987PhDT.......199B}, as implemented in the ROTCUR task within 
the NEMO software package \citep{1995ASPC...77..398T} (hereafter P3). 
The galactic disk is subdivided into rings, each of which is described 
by the usual 6 parameters. Starting with the initial estimate for the 
fitting parameters, these are then adjusted iteratively for each ring 
independently until convergence is achieved. We run NEMO considering 
or neglecting the variations of the orbital centers and systemic 
velocity for each ring.  The ROTCUR task gives less accurate results
than our P1 method since it does not take into account 
the distance variations between the observer and the far/near side 
of the galaxy and minimizes the free parameters of one ring at a time 
without subsequent iterations.  We ran ROTCUR with 24 free rings from 
8 to 34~kpc. For the last ring the solutions did not converge. We show in 
Figure~\ref{ring3} the resulting $i$, $\theta$ and $V_{sys}$ for 3 
cases. In one case the fit is done over all data points with uniform 
weight, while varying $V_{sys}$ and the orbital centers $x_c$ and $y_c$. In a 
second attempt, we fitted the data weighted with the cosine function 
of the galactic angle away from the major axis and excluding data 
within a 20$^\circ$ angle around minor axis.  In a third attempt, we 
fitted the data without considering possible orbital center shifts and 
setting $V_{sys}$ to the average value found by ROTCUR when $V_{sys}$  
was allowed to vary. This method confirms that the average systemic
velocity of the disk of Andromeda is slightly more negative than
previously thought and that there is a warp in outer disk which
brings the orbits closer to be edge-on.

We conclude that the 3 fitting  methods largely produce similar results
and will make use of the results of most accurate procedure, P1, in the rest 
of the paper.

\subsection{Comparison with previous results} 

As we mentioned in the Introduction, only recently  
a 21-cm survey made with a synthesis telescope has been used to model 
the warp and the rotation curve of M31 simultaneously 
\citep{2009ApJ...705.1395C}. The conclusion is similar
to ours, namely that the warp of M31 is such that the outer disk is at 
higher inclination and lower position angle than the inner disk. However a
close inspection reveals
some differences. First, the average value of the disk inclination between
10 and 28~kpc that \citet{2009ApJ...705.1395C} find is 74$^\circ$ while 
what we find is 77$^\circ$, more consistent with that derived from optical 
surface photometry \citet{1987A&AS...69..311W}.  Secondly, the
inclination of the outermost fitted radius is higher for our model compared to what
\citet{2009ApJ...705.1395C} find, while the position angle is slightly lower.
We list below three main differences in the two fitting methods which can help 
explaining the variations in the resulting tilted ring model parameters.

-We model the bulk rotational velocity of M31 using the mean rotational velocity 
observed at each position.
That is, we use the first moment map made from the 120~arcsec by 2~km~s$^{-1}$  
data cube, following 4-$\sigma$ blanking in the  
120~arcsec by 18.5 km~s$^{-1}$ data cube to determine what is included in the  
integral. \citet{2009ApJ...705.1395C} use the peak velocity and if there are
multiple components they use the peak velocity of the `main component'. 
The main component is defined
to be the one which has the largest velocity relative to the systemic velocity 
of the galaxy. Their choice is based on the observational evidence of multiple
peak profiles, very prominent in the inner regions. As we explain fully in
the next subsection we did not include the innermost region 
(from 0 to 8~kpc in radius) in our analysis because it is dominated by 
non-circular motions.
Even though taking the mean velocity we might be systematically biased to 
lower apparent rotational velocities, the bias effects  
are minimal at large radii where we are most interested to the dynamics.
In general, we find that assuming the peak velocity of the main component
as the best approximation to the rotation curve, as in \citet{2009ApJ...705.1395C},
is very model dependent. Moreover, to determine it in a 
robust way one has to assume certain observational conditions 
(i.e. to which level is a high velocity feature accepted in terms 
of signal-to-noise; also, what happens if the highest velocity 
feature is blended inside a bright component so
that only a `shoulder' but not a secondary peak can be seen in the spectra).
Even though it is true that observations
carried out with a low resolution and projection effects bias the mean 
velocity towards lower velocities, we feel that 
results might be more robust when considering the
mean velocity than  other estimates if the system is as complex, asymmetric   
and disordered as M31 (and observed with a high spatial resolution).
Accretion events, which M31 is experiencing, produce several distinctive 
morphological and dynamical 
signatures in the disk, including long-lived ring-like features, significant 
flares, bars and faint filamentary structures above the disk plane. 
In M31, the likely 
non-negligible disk thickness coupled with a complex, asymmetric  
warping and with the presence of non-circular motion related to
multiple spiral arm segments intersecting along the line 
of sight, makes difficult to assess the reliability of velocity indicators 
different than the mean for tracing the bulk circular motion of the disk.
To prove the complexity of the system,  we would like to point out
that  across the disk of M31 and especially along the bright ring-like
structure at 10-15~kpc in radius, double peak profiles
are often present. Sometime peaks are separated by more than 100~km~s$^{-1}$.
There is however not a systematic azimuthal or
radial pattern as to whether is the fainter or the brighter peak to show the 
most extreme velocity from systemic. If it is just the warped outer disk
superimposed with the main disk to cause the multiple peak profiles we
should have found a more systematic behavior since the neutral hydrogen
surface density decreases considerably beyond 15~kpc. 
High resolution IR maps of M31 \citep{2006ApJ...638L..87G}
show that the northern half the 15~kpc  
arm, distinct on the major axis, merges into other 
spiral arms (or ring like structure) at 10 kpc. Peak velocities might more 
closely trace non-circular motion of the arms and produce wiggles in the rotation
curve which do not average out when additional perturbations are
present. This is the case for M31, in which  the southern
half is more strongly tidally perturbed than the northern half.  Such
a curve cannot be reproduced without modelling the spiral arms locally
in term of mass condensation and non-circular motion.

-As we mentioned earlier in this Section we believe that it is relevant for
M31, a very extended and nearby galaxy, to derive galactocentric distances
in the frame of a tilted ring model using appropriate spherical geometry. This
takes into account the fact that the near side of the galaxy is effectively 
closer than the far side of the galaxy. Procedures often available for
deriving the kinematical parameters are built for the more numerous more
distant galaxies and do not account for this effect.
Moreover, our routine works with 66 free parameters simultaneously since
the best fitting values of the parameters for each ring  cannot be found independently 
from those relative to other rings when a warp is present. In fact the warp makes 
two or more rings overlap along the line of sight and in the overlapping region 
the expected velocity depend on all parameters of the rings which overlap.

-In fitting the tilted ring model to the data we do not apply any angular
dependent weight i.e. the same weight is given to pixels close to minor
axis than to major axis. This is because we would like to minimize the
risk of amplifying the kinematical signatures of sporadic features which 
happens to lie close to major axis. As we will see in the next Section,
our model produces a stable rotation curve, not very sensitive to the
choice of the opening angle $\alpha_{max}$ around the major axis.

\subsection{Why exclude the innermost region from fitting} 

Opposite to \citet{2009ApJ...705.1395C} we do not include the inner
regions in our analysis. This is because
after the pioneer work of \citet{1956StoAn..19....2L}
several other papers have pointed out in the inner region of M31 the presence of 
morphological structures, such as a bar and a bulge, associated with streaming 
motion and non-circular orbits 
\citep[e.g. ][]{1994ApJ...426L..31S,2006MNRAS.370.1499A,2006ApJ...638L..87G,
2007ApJ...658L..91B}. In particular \citet{2001A&A...371..476B,2002MNRAS.336..477B}
have shown that the anomalous velocities observed in the inner region of M31
can be explained as the response of the gas to the
potential of a triaxial rotating bulge. Using a bulge effective radius of 
10~arcmin they have derived which family of periodic elliptical orbits 
exist. They find that the bulge gives a non-negligible
contribution to the galaxy potential out to about 7~kpc, and only
at larger radii circular motion related to the disk gravitational potential
dominates. The model well
reproduced the velocities observed through the CO J=1-0 line emission.
Since in this paper we are only modelling the large scale circular motion of 
the disk we use the mean velocity of the HI gas as tracer of the circular
velocity from 8~kpc outwards.


\section{The rotation curve and the radial distribution of the baryons}

\begin{figure} 
\includegraphics[width=\columnwidth]{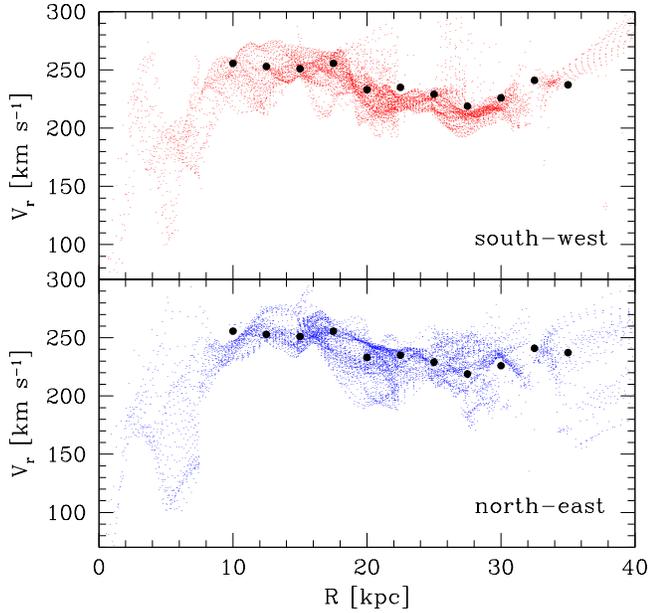} 
\caption{Rotational velocities derived for the northern and southern 
halves of M31 using the geometrical parameters of the best fitting tilted 
ring model as shown in Figure~\ref{ring1}. For this Figure 
$\alpha_{max}=30^\circ $. The filled circles indicate the best fit rotational 
velocity parameters derived by the tilted ring model fit P1.} 
\label{rot1} 
\end{figure}

\begin{figure} 
\includegraphics[width=\columnwidth]{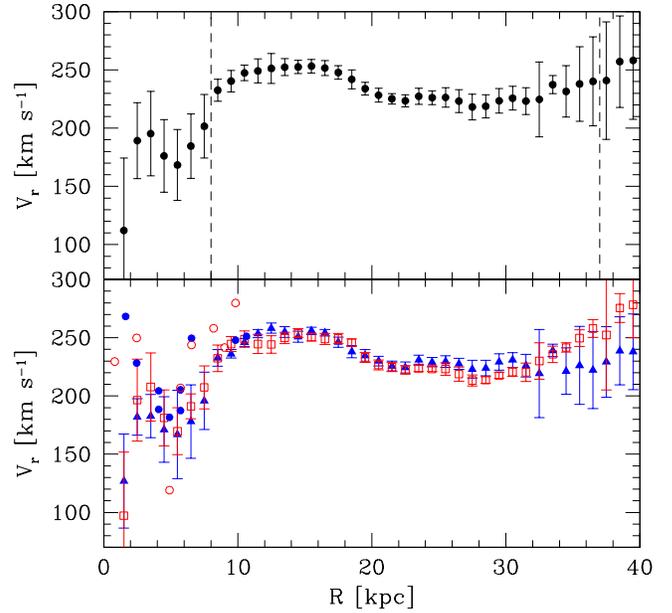} 
\caption{The bottom panel shows the rotation curves obtained for the 
northern and southern halves of the galaxy by averaging the data shown 
in Figure~\ref{rot1} in radial bins 1~kpc wide. Errorbars refer to the 
dispersion around the mean. Filled triangles are 
for the northern half, open squares for 
the southern half of the galaxy.  
Rotational velocities in the inner region derived from CO data  
by \citet{1995A&A...301...68L} are shown with filled and open circles 
for the northern and southern major axis, assumed to be at PA=38$^\circ$ 
and with $i=77^\circ$. The filled dots in the upper panel show  
the global rotation curve. As explained in the text, errorbars take into 
account the dispersion around the mean and the differences between the  
values for the two halves of the galaxy. 
The vertical dashed line mark the innermost radius 
which will be considered in this paper for the dynamical analysis. 
} 
\label{rot2} 
\end{figure}

We now apply the geometrical parameters of the best fitting tilted ring model 
to derive the rotation curve of the galaxy. We set the inclination, 
position angle and systemic velocity to the values shown by the heavy 
continuous line in Figure~\ref{ring1}, and consider also the small shifts  
of the orbital centers obtained by our minimization procedure P1.  
We derive the rotation curve by averaging the rotational velocities of 
data points in radial bins 1~kpc wide. For radii 
between 8 and 10~kpc, we extrapolate 
the model parameters of the innermost ring centered at 10~kpc. For radii larger  
than 35~kpc, we extrapolate the model parameters of the outermost ring  
at 35~kpc. Just for curiosity, we also checked what would be the rotation curve of
the inner regions for our dataset if we assume circular motion and
inside 8~kpc we set the disk inclination equal to the inclination 
of the ring at 10~kpc and we consider 28$^\circ$ as position angle   
(this is the average value we derive from an inspection of the mom-1 map). For  
this innermost region we consider zero shifts for the orbit centers and 
-306~km~s$^{-1}$ as systemic velocity. Figure~\ref{rot1} and Figure~\ref{rot2}
show the large dispersion in the velocities relative the central region
due to the presence of multiple components and to the uncertainties related 
to orbital eccentricities inside 8~kpc, discussed in the previous Section.
To complement 21-cm data in the inner regions we show in Figure~\ref{rot2} 
the peak brightness velocities of CO lines \citep{1995A&A...301...68L}. 
These have been observed along the major axis assumed to be at PA=38$^\circ$ 
and with $i=77^\circ$. However, notice that this is shown not just
to point out the consistency of the molecular and atomic gas 
velocities, but to emphasize one has to consider non-circular motion to 
properly trace the rotation curve in the inner region \citep{2001A&A...371..476B,
2002MNRAS.336..477B}.  Hence we shall not use the CO data as well as
the 21-cm data at radii smaller than 8~kpc. 
In the rest of the paper, we will only analyze the rotation curve 
between 8 and 37~kpc.  Beyond 37~kpc the northern and southern
halves do not give consistent rotation curves for any of the deconvolution
models. This is likely due to highly perturbed orbits in the outermost
regions, especially for the southern half which is closer to M32. 

We consider points  
which lie within an opening angle $\alpha_{max}$ on either side of the major axis.  
We first derive the rotation curve in the northern and 
southern halves separately. We check the consistency of the rotation curves 
for different values of $\alpha_{max}$, with the value of the rotational  
velocities determined by the tilted ring model over the whole galaxy,  
and between the northern and the southern halves. 
When we vary  $\alpha_{max}$ between 15$^\circ$ and 75$^\circ$, we obtain 
the same rotation curves consistently (variations are less than  
1~km~s$^{-1}$).  Only the dispersion around the mean 
increases slightly as we increase $\alpha_{max}$. We shall use   
$\alpha_{max}=30^{\circ}$ for the rest of the paper. 
The rotation velocities in the two halves are consistent (within 3-$\sigma$).  
At many radii, mean velocities in the northern and southern halves 
differ by less than  1-$\sigma$ corresponding to  
only a few km~s$^{-1}$ 
at most, as shown in Figures~\ref{rot1} and~\ref{rot2}. 
In Figure~\ref{rot1} we also  
display the rotational velocity parameter of the best tilted ring model.    
The best fitting ring model gives the most consistent rotation curves.  
We also tried to deconvolve the observed velocities using other tilted  
ring models shown in Figure~\ref{ring1}, but they give less consistent  
rotation curves. Average velocities are derived in the two halves by  
assigning a weight to each pixel equal to the integrated brightness intensity 
i.e. to the HI column density along the line of sight in the limit of 
optically thin 21-cm line. The global rotation curve is the arithmetic 
mean of the average rotational velocities in the two galaxy halves. 
The errorbars of the global rotation curve shown in the upper panel 
of Figure~\ref{rot2} are computed as 

\begin{equation} 
\sigma_g={V_{hi}-V_{low} \over 2} + {\sigma_{hi}+\sigma_{low}\over 2} 
\end{equation} 

\noindent 
where $V_{hi}$ refers to the highest rotational velocity and $V_{low}$ to the 
lowest rotational velocity between the two mean velocities relative   
to the northern and to the southern galaxy side; $\sigma_{hi}$ and $\sigma_{low}$ 
refer to the relative dispersions around the mean. 

In Table 1 we give the parameters of the best fitting tilted ring model
and of the average rotation curve of M31 in the radial interval which
will be used in the next Section for the dynamical mass model.
The position shifts of the orbital centers are rather small and
can be neglected. The systemic velocity shifts,$\Delta$V$_{sys}$,
are computed respect to the nominal value V$_{sys}$=-300~km~s$^{-1}$.

\subsection{A comparison with other rotation curve measures}

\begin{figure} 
\includegraphics[width=10cm]{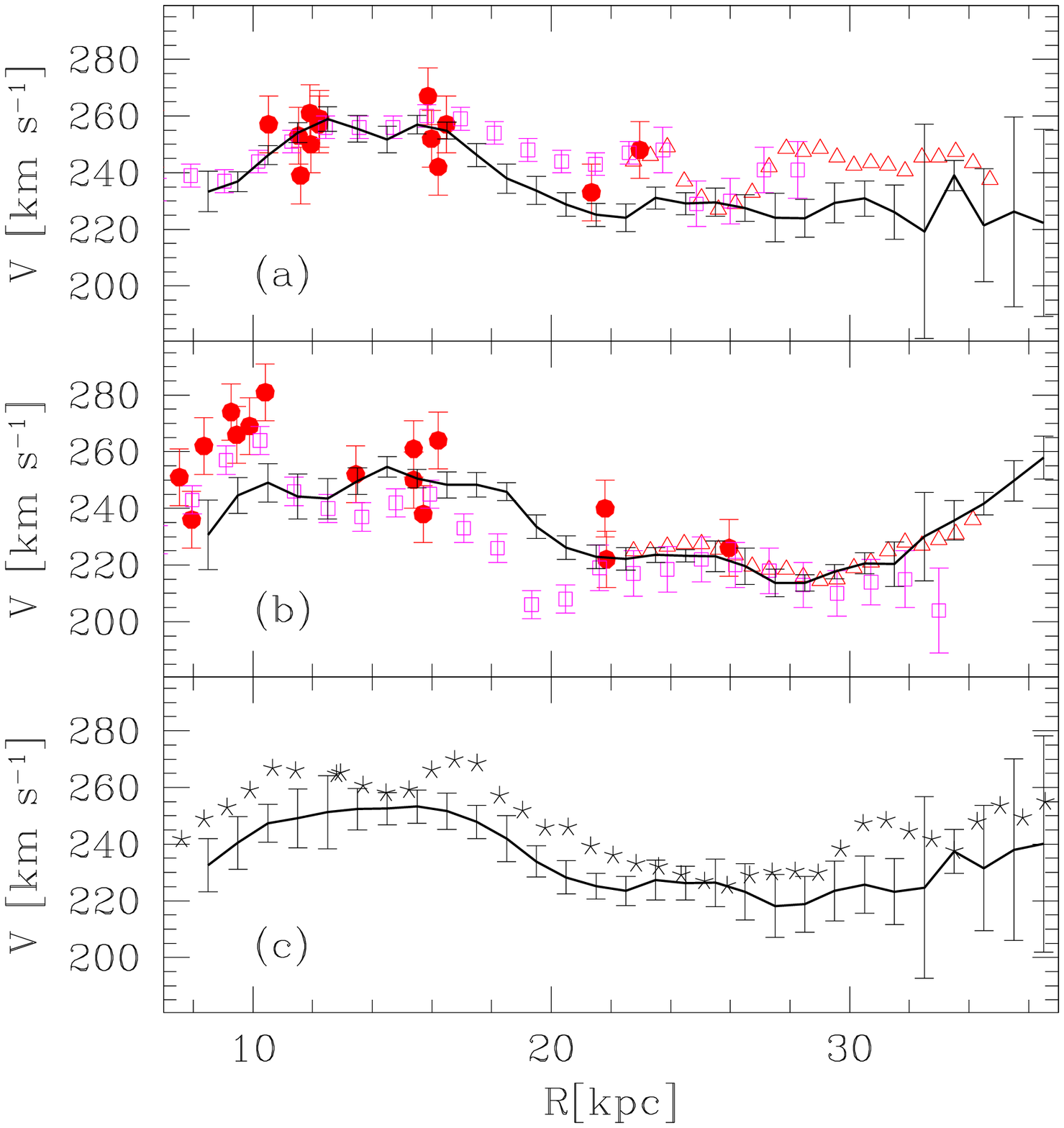} 
\caption{ The rotation curves derived in this
paper (continuous line) and some previously determined ones. In panel
$(a)$ we display data for north-east galaxy side, in panel $(b)$ for
the south-west side and in panel $(c)$ the average values. Filled
circles are for \citet{1989PASP..101..489K} optical data along the 
major axis,
open squares and open triangles are for \citet{1976MNRAS.176..321E} 
and \citet{1977MNRAS.181..573N} 21-cm data along the
major axis. Asterisk symbols are used in $(c)$ to display the 
average rotation curve of  \citet{2009ApJ...705.1395C}. Original
data in $(a)$ and $(b)$ have been scaled according to an
assumed distance $D=785$~kpc and systemic velocity 
$V_{sys}=-306$~km~s$^{-1}$. 
} 
\label{compa} 
\end{figure}

In Figure~\ref{compa} we show for comparison the rotation curves 
derived in this paper (continuous line) and some previously determined 
ones. In panel
$(a)$ we display data for north-east galaxy side, in panel $(b)$ for
the south-west side and in panel $(c)$ the average values. 
Original data in $(a)$ and $(b)$ have been scaled according to an
assumed distance $D=785$~kpc and systemic velocity 
$V_{sys}=-306$~km~s$^{-1}$. 
We show both optical data from  \citet{1989PASP..101..489K}
and 21-cm data from \citet{1976MNRAS.176..321E,1977MNRAS.181..573N}.
Unfortunately most of the previous determinations rely on
an assumed fixed inclination for the disk and on data along
the major axis alone. That implies that possible local velocity 
perturbations will not be averaged out. These are clearly visible
especially in the literature data between 8 and 20~kpc 
relative to the south-west galaxy half
(panel $(b)$), strongly perturbed by the M32 tidal interaction.
In the northern side we derive a somewhat lower rotational
velocity, perhaps due to the presence of the warp which implies a higher
disk inclination, not accounted for by previous data analysis.
Taking into account these limitations, the general 
agreement seems good.
The asterisk symbols are used in $(c)$ to display the 
average rotation curve of  \citet{2009ApJ...705.1395C} which
lies above ours, due especially to somewhat lower inclination the 
authors derive for the tilted ring model. Also, despite their
 analysis masking some perturbations such as the
external arm, their curve is less smooth than ours. As discussed already
in the previous Section, this might
be due to different choices of velocity components to extract
the rotation curve or to their use of weighting function which
gives more weight to data points close to major axis. A side
effect of this choice is to  
retain any local velocity perturbation present along major axis.

\begin{figure}
    \centering  
\includegraphics[width=10 cm]{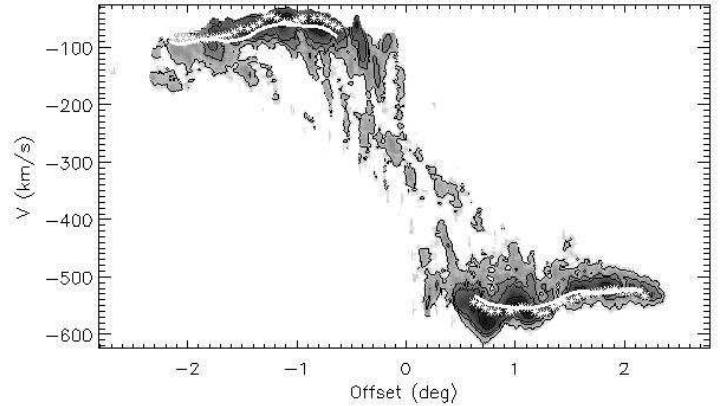} 
\caption{ Position velocity diagram along the photometric major axis,
at a position angle of 38~$^\circ$. A logarithmic scale is used for
the brightness and its contours are drawn at 0.013,0.06,0.13,0.6 Jy/beam.
The lowest contour is at the 3-$\sigma$ level since $\sigma$ is 4.2 mJy/beam.
The continuous line show the rotation curve adopted in this paper while the
asterisk trace the rotation curve of \citet{2009ApJ...705.1395C}.
In both cases, the rotational velocities have been corrected using
the average value of the disk inclination and systemic velocity
found by the authors in the region shown.
} 
\label{pv} 
\end{figure}

A position velocity plot made along the photometric major axis,
at a position angle of 38~$^\circ$, is shown in Figure~\ref{pv}.
Our adopted average rotation curve, projected back along major axis,
has been superimposed to it (diamonds).
The average value of the disk inclination and systemic velocity,
as derived in our best fitting tilted ring model in the radial interval 
of interest, has been used. For a comparison
we display also the average rotation curve of \citet{2009ApJ...705.1395C}
after applying the disk inclination and systemic velocity
corrections adopted by the authors (asterisk symbols). The figure clearly 
shows that 
despite the different gas velocities adopted by the two teams to trace the 
rotation curve, as explained in detail 
in Section 2, differences in the rotational velocities at large radii
before deprojection are marginal. 
The difference between the two measurements becomes
more significant when rotation curves (deprojected) are directly compared. 
This illustrates the relevance of the differences in the 
parameters of the best tilted ring model found 
by the two teams. In particular \citet{2009ApJ...705.1395C} derive lower
inclination angles for the M31 disk and hence higher rotational
velocities. The agreement between the PV diagram
along the major axis and the component of the rotational velocities 
measured along the line
of sight improves, as it should, when using only data in 
a very small opening angle around the major axis and data in the two
galaxy halves are kept separate. Notice, for example, the higher
velocity present along the major axis around 0.7 angular offset
respect to the average rotation curve. That is also visible 
in the optical data shown in the middle panel of Figure~\ref{compa}.
This anomalous velocity present along the major axis is averaged
out when data from other directions is taken into account.

\begin{table}
\caption{ The HI rotation curve of M31 and the parameters of the best fitting 
tilted ring model}
\label{rotcur}
\begin{tabular}{lcccccccc}
\hline \hline
    R&  i& $\theta$& $\Delta$x$_c$& $\Delta$y$_c$& $\Delta$V$_{sys}$& V$_r$& 
    $\sigma$(V$_r$) \\
     (kpc)&  ($^\circ$)& ($^\circ$)&  ($^\prime$)& ($^\prime$)& (km~s$^{-1}$)& 
     (km~s$^{-1}$)& (km~s$^{-1}$) \\
\hline \hline
       8.5 &  77.9 & 37.7 & -1.6 &  2.3 &  -10.0 &  232.0 &  11.0 \\
       9.5 &  77.9 & 37.7 & -1.6 &  2.3 &  -10.0 &  240.7 &   8.8 \\
      10.5 &  77.9 & 37.7 & -1.5 &  2.3 &   -9.4 &  247.6 &   6.5 \\
      11.5 &  78.0 & 37.6 & -1.4 &  2.2 &   -8.3 &  249.1 &  10.7 \\
      12.5 &  78.1 & 37.6 & -1.3 &  2.2 &   -7.2 &  251.2 &  13.5 \\
      13.5 &  78.3 & 37.1 & -1.0 &  1.3 &   -3.4 &  252.6 &   7.6 \\
      14.5 &  78.6 & 36.6 & -0.7 &  0.3 &    0.4 &  253.2 &   5.6 \\
      15.5 &  78.3 & 36.3 & -0.9 &  0.2 &    0.1 &  253.8 &   6.5 \\
      16.5 &  77.5 & 36.2 & -1.8 &  0.8 &   -4.1 &  251.6 &   7.1 \\
      17.5 &  76.7 & 36.1 & -2.6 &  1.4 &   -8.4 &  247.3 &   5.2 \\
      18.5 &  76.6 & 36.4 & -1.7 &  0.8 &   -7.4 &  241.9 &   8.1 \\
      19.5 &  76.5 & 36.7 & -0.9 &  0.2 &   -6.5 &  233.6 &   4.6 \\
      20.5 &  76.2 & 37.0 & -0.3 &  0.0 &   -6.0 &  227.4 &   5.5 \\
      21.5 &  75.7 & 37.5 &  0.1 &  0.3 &   -6.0 &  224.0 &   5.3 \\
      22.5 &  75.1 & 38.0 &  0.5 &  0.5 &   -6.0 &  223.2 &   5.4 \\
      23.5 &  76.1 & 38.3 &  0.4 &  0.4 &   -6.0 &  227.3 &   7.0 \\
      24.5 &  77.0 & 38.6 &  0.4 &  0.4 &   -6.0 &  226.2 &   6.0 \\
      25.5 &  77.7 & 38.3 &  0.3 &  0.2 &   -6.0 &  226.4 &   8.4 \\
      26.5 &  78.0 & 37.6 &  0.3 & -0.1 &   -6.0 &  223.6 &   9.5 \\
      27.5 &  78.3 & 36.9 &  0.3 & -0.3 &   -6.0 &  218.9 &  12.0 \\
      28.5 &  79.3 & 35.7 &  0.1 & -0.3 &   -6.0 &  218.8 &   9.9 \\
      29.5 &  80.2 & 34.5 & -0.1 & -0.3 &   -6.0 &  223.5 &  10.6 \\
      30.5 &  81.6 & 33.0 & -0.1 & -0.3 &   -6.0 &  225.8 &  10.1 \\
      31.5 &  83.5 & 31.3 &  0.0 & -0.2 &   -6.0 &  223.2 &  11.6 \\
      32.5 &  85.3 & 29.6 &  0.2 & -0.2 &   -6.0 &  224.7 &  32.1 \\
      33.5 &  85.6 & 30.9 &  0.2 &  0.0 &   -5.7 &  237.4 &   7.8 \\
      34.5 &  85.9 & 32.2 &  0.3 &  0.1 &   -5.4 &  231.5 &  22.1 \\
      35.5 &  86.0 & 32.9 &  0.3 &  0.2 &   -5.2 &  238.0 &  32.0 \\
      36.5 &  86.0 & 32.9 &  0.3 &  0.2 &   -5.2 &  240.1 &  38.2 \\

\hline \hline
\end{tabular}
\end{table}

\subsection{The bulge-disk decomposition and the stellar distribution} 

Radial profiles of stellar disk surface brightness can usually be represented  
by an exponential distribution. The mass-to-light ratio is 
determined by the dynamical analysis of the rotation curve or by the optical  
colors once the radial exponential disk scale length is known.  
The disk scale length can be established from surface brightness profiles in  
various bands and it might be wavelength dependent. In Andromeda, it varies from  
7.7~kpc in the U-band to 5.9 in the R-band \citep{1988A&A...198...61W} to 4.5~kpc  
in the K-band \citep{1983PASJ...35..413H, 
1986A&A...161...70B}. The images used at optical wavelengths  
\citep{1988A&A...198...61W} can trace the disk surface brightness out to  
galactocentric radii of about 25~kpc, 
while the available K-band images (including that obtained through the 2MASS 
survey) loose sensitivity beyond 20~kpc. The discrepancy 
between the optical and infrared scalelength can be easily explained in terms 
of a radially varying star formation history or of an extinction 
gradient. Light in the K-band traces the 
mass distribution better than optical light because of the reduced extinction 
and because most of the stellar mass in galaxy disks is due to old, low mass  
stars. Therefore for the dynamical analysis, K-band scale lengths are usually 
preferred despite the reduced sensitivity of infrared images. 
Recent mid-infrared observations of Andromeda obtained with the 
Infrared Array Camera on board of the Spitzer Space Telescope  
\citep{2006ApJ...650L..45B} show a disk scale length of 6.1~kpc at 3.6~$\mu$m 
measured on the same radial interval as measured by \citet{1986A&A...161...70B} 
using the K-band light. 
A larger scale length at 3.6~$\mu$m is also found in M33 by comparing 
the Spitzer image with the 2MASS K-band image \citep{2009A&A...493..453V}.  
This might imply that intermediate age, cool supergiants contribute a 
substantial fraction of the NIR emission at 3.6~$\mu$m \citep{2008ApJ...687..230M}.

The determination of the disk scale length depends also on the bulge-disk 
light decomposition and K-band images might not be deep enough to obtain an 
accurate bulge-disk decomposition. Nevertheless we feel that the shorter disk 
scalelength measured in the K-band images compared to 3.6~$\mu$m images
is not affected by this limitation. 
M33 in fact has no bulge and indeed the scale length at 2.2~$\mu$m is shorter than 
at 3.6~$\mu$m.  
A spherical model for the bulge of M31 is a simplification. Detailed modelling of  
the surface brightness shows that at very least the bulge is an oblate spheroid with 
axis ratio of 0.8 \citep{1983ApJ...266..562K} but most likely it is a triaxial 
bulge \citep[e.g. ][and references therein]{2001A&A...371..476B}.  
For the purpose of this paper, which is the dynamical mass modelling 
of the disk rotational velocities (beyond 8~kpc), we can neglect the  
asphericity of the bulge which affects the orbits in the innermost few kpc.  
Using a R$^{1/4}$ de Vaucouleurs law (i.e. a Sersic index of 4) for the light 
distribution it is customary to characterize the bulge by its  
effective radius which encloses half of its total light. 
At optical wavelengths (in the R- and V-band) \citet{1988A&A...198...61W}  
found a bulge effective radius of 2.3~kpc and a bulge to disk luminosity  
ratio B/D=0.82. Using the same dataset in the R-band 
combined with Hubble Space Telescope data of \citet{1993AJ....106.1436L} 
and data from \citet{1983ApJ...266..562K} for the innermost regions 
\citet{2006MNRAS.366..996G} derived a much smaller bulge scale radius: 
0.61~kpc (similar to what can be inferred by inspecting the 2MASS images).  
Using the 3.6~$\mu$m Spitzer image  
\citet{2006ApJ...650L..45B} modelled the bulge light distribution with a 
R$^{1/4}$ de Vaucouleurs law and found a 1.7~kpc effective radius and a bulge 
to disk light ratio of B/D=0.78. Using the same data \citet{2008MNRAS.389.1911S} 
obtain a bulge effective radius of 1.93~kpc for a Sersic index n=1.7, 
and a disk scalelength of 5.91~kpc 
with B/D=0.57. Fits to the bulge light distribution using smaller Sersic  
indexes have also been done by \citet{2005ApJ...631..820W} using a 
brightness profile from an 
I-band image out to 24~kpc: for a Sersic model with n=1.6 the bulge effective  
radius was found to be 0.89~kpc and the disk scale length 5.7~kpc, even though 
a steepening of the scale length is clearly visible beyond 15~kpc. 
An faint stellar disk which extends as far as 40~kpc from the Andromeda 
nucleus has been recently pointed out  by \citet{2005ApJ...634..287I} with  
an exponential scale length of 5.1~kpc in the I-band. 

Given the uncertainties in the disk and bulge mass distribution we 
will attempt to fit the rotation curve using 
a varying disk 
scalelength between 4.5 and 6.1~kpc in steps of 0.2~kpc.
For the bulge we shall use 4 possible parameter combinations, namely an 
effective radius of 2.0 and 0.7~kpc and a Sersic index n=4, and n=1.6. 
Given the fact that we cannot constrain the dynamical contribution of the
bulge since we are not fitting the motion in the inner regions, our purpose
will be only to see if the
dynamical fit to the disk improves considerably when 
using any of the four combinations.

\subsection{Stellar mass-to-light ratios} 

Analysis of the star formation histories of the bulge and disk of M31 suggest 
that there is no age difference between the bulge and the disk  
\citep{2006AJ....132..271O}. However previous attempts to fit former rotation 
curves of this galaxy found a higher mass-to-light ratio for the disk than for the 
bulge \citep{2003ApJ...588..311W}, which was unexpected given the older age 
of bulges relative to disks.  
Hydrodynamic simulations of the triaxial bulge of M31 by  
\citet{2001A&A...371..476B} found a B-band mass-to-light ratio of 6.5 
for the bulge i.e. a stellar mass of $10^{10}$~M$_\odot$. 
In the disk of Andromeda there is also a color 
gradient visible in the disk \citep{1988A&A...198...61W,2008MNRAS.389.1911S}  
since B-R varies between 
1.7 in the inner regions to 1.3 in the outer regions  (this can be due to 
changes in  metallicity, age or  extinction). According  
to \citet{2001ApJ...550..212B} the mass-to-light ratio in the K-band expected  
from B-R colors can vary from the value of 1 in the central regions to 0.65 in  
the outer disk if extinction does not change radially. 
The mass-to-light ratio in the B-band can vary between 8 and  
2.5~M$_\odot/$L$_\odot$ and we will consider these two values as the extreme  
acceptable disk mass-to-light ratio values. Since we fit the rotation curve beyond 
8~kpc we cannot constrain the bulge mass-to-light ratio and hence we will 
consider also models with equal mass-to-light ratios for the bulge and the 
stellar disk. 

The bulge and disk blue luminosities which we shall use
to compute the mass-to-luminosity ratio are $9\times 10^9$~$L_\odot$ 
and $2.1\times 10^{10}$~$L_\odot$ respectively. 
These have been derived
using the integrated B-band magnitude measured by \citet{1988A&A...198...61W},
corrected for absorption, assuming that the bulge contribution is 30~$\%$ of
the total emission in the B-band. As we mentioned in the previous subsection, 
the decomposition
of the light profile into a bulge and disk component is somewhat uncertain
and especially the bulge integrated luminosity is not firmly estabished yet
\citep[see also ][]{1989AJ.....97.1614K}.

\subsection{The gas surface density} 

\begin{figure} 
\includegraphics[width=\columnwidth]{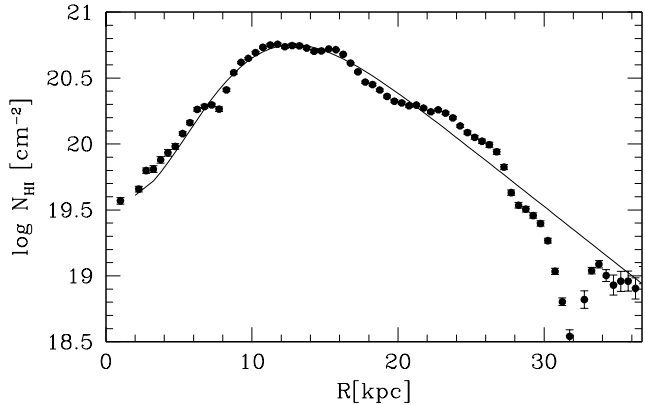} 
\caption{The neutral atomic hydrogen column density perpendicular to 
the galactic plane is shown as a function of galactocentric distance. The 
best fitting tilted ring model P1 is used for the deconvolution of the 
21-cm line brightness image. The optically thin line approximation has 
been used to convert the surface brightness in HI gas column density. 
The continuous line shows the total atomic gas surface density of the M31 
disk as modelled for the dynamical analysis.} 
\label{himass} 
\end{figure}

For the gaseous disk we shall consider the atomic hydrogen 
and the molecular gas surface density. These will be multiplied   
by 1.33 to account for the presence of helium. 
Using the best fitting tilted ring model (P1) we derive the   
radial distribution of neutral atomic gas, perpendicular to the galactic plane,   
in the optically thin approximation. This is shown in Figure~\ref{himass}  
as a function of the 
galactocentric radius. The corresponding total HI mass is $5.4\times 10^9$~M$_\odot$. 
To consider the possible presence of opaque 
clumps we can multiply the HI surface brightness by 1.3 since 
this is the correction inferred by \citet{2009ApJ...695..937B}. This correction 
has however no noticeable effect for the dynamical analysis carried out in the 
next Section. 

The continuous line in Figure~\ref{himass} is the
log of the function f$_{HI}(R)$.   
We shall use f$_{HI}(R)$ to compute the dynamical contribution of the 
atomic gas mass to the rotation curve. It has the following expression

\begin{equation}
f_{HI}(R) = e^{2.3(22.0888-0.08759\ R)\ exp(-0.11773\ 
          exp(-0.03276\ R^{1.7}))}
\end{equation}

and it provides a good fit to the atomic surface
density distribution perpendicular to the galactic plane. 
The sharp decline of the HI beyond  
27-30~kpc is likely due to the ionization of the atomic hydrogen by the local  
extragalactic 
UV background radiation \citep{1993ApJ...419..104C}. Hence the fitting function 
which approximates the total atomic gas distribution is shallower than the 
HI distribution in the outer region. There is also ionized atomic gas  
in the inner regions, as can be inferred from the H$\alpha$ emission maps
\citep{1994ApJ...431..156W}. 
However quantifying the column density and the geometrical distribution of 
such ionized component is subject to several assumptions. 
The gaseous mass involved is small and irrelevant to the rotation curve 
fitting since in the inner region it is the stellar component which is 
dynamically dominant. Hence we shall neglect it. 

For the molecular gas fraction we considered the average radial 
variation (North+South) as shown in Figure~10 of  
\citet{2006A&A...453..459N}.  

\section{Dynamical analysis} 

In this Section we will analyze the mass distribution in the Andromeda 
galaxy. We first attempt to fit the rotation curve of Andromeda, 
derived in the previous Section, from 8 to 37~kpc without a dark matter 
halo. Then we carry on the dynamical analysis considering two different  
models for the radial distribution of the dark matter 
density distribution. We will discuss at the end 
the resulting total mass of Andromeda  in the context also of data at very  
large radii from observations of the Andromeda satellites and other objects. 
We shall consider both models in which the the disk and the bulge have
the same mass-to-light ratio 
and models in which these are  two independent variables. Unless stated
differently we shall consider the HI gas in the optically thin approximation. 
The half thickness of the disk is assumed to be 0.5~kpc  and the contribution
of the disk mass components to the rotation curve is computed according to
\citet{1983MNRAS.203..735C}. We use the reduced chi-square statistic,  
$\chi^2$, to judge the goodness of a model fit. 
We consider first 2 free  parameters: the disk and the bulge mass-to-light
ratio $(M/L)_{d,b}$,  
which we vary continuously. For the stellar disk we vary the 
exponential scalelength $h_d$ using steps of 0.2~kpc in the interval 
4.5$\le h_d \le 6.1$~kpc. For the bulge we consider 2 possible effective radii  
$h_b$ and Sersic index n=4 and n=1.7.

\subsection{Newtonian and non-Newtonian dynamics without dark matter} 

\begin{figure} 
\includegraphics[width=84mm]{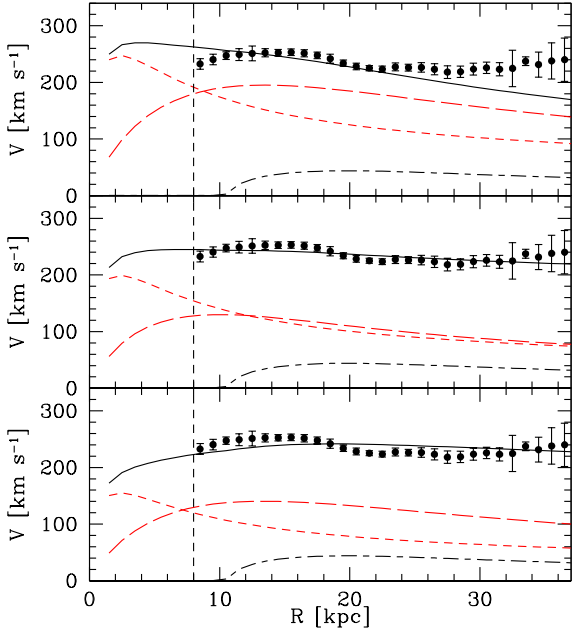} 
\caption{The 21-cm rotation curve for M31 and the 
best fitting mass model with no 
dark halo in the Newtonian case (top panel) and for MOND (middle panel). 
The short-dashed lines indicate the Newtonian stellar bulge and disc 
contribution to the rotation curve. The dot-dashed line is the gas contribution. 
The figure displays the the best fitting mass model which is obtained for  the
Newtonian case when
$(M/L)_d=(M/L)_b=8.0 \ M_\odot/L_\odot$, $h_d = 6.1$~kpc, $h_b = 2.0$~kpc and
n=1.6. The MOND best fit ($\chi^2=1.3$) is obtained
when the free parameters  are  $h_d=4.5$~kpc, $(M/L)_d=2.5 \ M_\odot/L_\odot$  
$(M/L)_b=5.2 \ M_\odot/L_\odot$, and $h_b=2$~kpc, n=1.6. 
For comparison we show in the bottom panel the MOND best fit when  
the disk scalelength $h_d$ is 6.1~kpc. The reduced $\chi^2$ in this case 
is higher ($\chi^2=3.1$), 
$(M/L)_d=4.1 \ M_\odot/L_\odot$, and $(M/L)_b=3.1 \ M_\odot/L_\odot$ .
} 
\label{mond_nodm} 
\end{figure}

We first attempt to fit the rotation curve in the framework of Newtonian 
dynamics without considering a dark matter halo. 
The best fit is obtained for a disk 
scalelength of 6.1~kpc with $(M/L)_d=(M/L)_b=8.0 \ M_\odot/L_\odot$, which 
is the maximum allowed value. The fit is slightly better when a bulge effective  
radius of 2~kpc is used. This is shown in the top panel of
Figure~\ref{mond_nodm}. However the fit is 
generally poor being the reduced $\chi^2>6$ for all possible combinations 
of parameters (see Table 2). Hence the model with no dark matter  
fails to fit the data under the assumption of Newtonian gravity. 
The fit stays poor even if unreasonably high values for the stellar mass-to-light 
ratio are considered ($>8 \ M_\odot/L_\odot$).
The evident failure is due to the declining rotation curve  
predicted by the baryonic mass distribution beyond 26~kpc. 

An alternative explanation for the mass discrepancy has been proposed by Milgrom   
by means of the modified Newtonian dynamics or MOND \citep{1983ApJ...270..365M}.  
Outside the bulk of the mass distribution, MOND predicts a much slower decrease of   
the (effective) gravitational potential, with respect to the Newtonian case. This    
is often sufficient to explain the observed non-keplerian behavior of RCs 
\citep{2002ARA&A..40..263S}. According to this theory the dynamics becomes   
non-Newtonian below a limiting  acceleration value, ${a_0}\sim 10^{-8}$~cm~s$^{-2}$,  
where the effective gravitational acceleration takes the value  
$g = g_n/\mu(x)$, with ${g_n}$ the acceleration in Newtonian dynamics
and $x=g/a_0$. Here we shall use the critical acceleration value $a_0$    
derived from the analysis of a sample of rotation curves   
$a_0 = 1.2\times 10^{-8}$~cm~s$^{-2}$ \citep{2002ARA&A..40..263S}.
We have tested MOND for different choices of the interpolating function $\mu(x)$
\citep[see ][ for details]{2005MNRAS.363..603F}. In particular we have used the
`standard' and the `simple' interpolation function and found that the `simple'
interpolation function provides slightly better fits to the M31 data. Hence
we shall use the `simple' interpolating function $\mu(x)=x/(1+x)$ in the rest of
the paper.

\citet{2007MNRAS.374.1051C} have already tested the former M31 rotation 
curve in the MOND framework. They concluded that, in M31, MOND fails  
to fit the falling part of the rotation curve at intermediate radii. 
However, this assessment was made using lower quality data and in the absence
of an appropriate knowledge of the tilted ring model parameters. 

Figure~\ref{mond_nodm} shows the MOND best-fit model curve when  
$(M/L)_d$ and $(M/L)_b$ are two independent free parameters (bottom panel). The 
mass-to-light ratio best fitting  
values are $(M/L)_d=2.5$, $(M/L)_b=5.2$, and $h_d=4.5$~kpc. The value of 
$\chi^2$ is 1.3, hence MOND provide a good fit to the data.
If we reduce the number of free parameters to just one by setting  
$(M/L)_d=(M/L)_b$ 
the fit is still reasonably good: the lowest $\chi^2$ is 1.37 corresponding to
$(M/L)_d=(M/L)_b=3.3 \ M_\odot/L_\odot$. The rotational velocities  
predicted by MOND are only slightly higher than observed for $20\le R\le 30$~kpc 
and slightly lower than the data for $10\le R\le 20$~kpc. 
As shown in Table 2, the goodness of the fits are not very sensible to the 
bulge mass-to-light ratio, since we are excluding the central regions. It is
however to be noticed that MOND best fits to the actual data require a higher 
mass-to-light ratio for the bulge than for the disk, in agreement with the 
expected older age of the bulge stellar component. 
In M31 it is around
10~kpc that $g_n\sim a_0$ and non-Newtonian corrections start to be important and  
force a falling Newtonian RC into a flat one, more consistent with the data. 
We notice that for MOND the value of the disk scalelength used to fit the
baryonic matter distribution is very important. In fact the fit  
becomes  poor if $h_d=6.1$~kpc, as shown in Figure~\ref{mond_nodm}. 
If future photometric studies will
confirm a disk scalelength of 6.1~kpc then one will have to consider 
possible variations of the assumed distance to M31 to make MOND predictions
more consistent with the kinematics traced by 21-cm data.

\subsection{Dark matter halo models} 

In the previous subsection we have seen that Newtonian dynamic fits to 
the rotation curve  without considering a dark matter halo are rather poor. 
We will now use the M31 rotation curve presented in this paper  
to test in detail the consistency of a possible halo density profile  
with theoretical models which predict a well defined dark matter  
distribution around galaxies. Namely, we 
shall consider a spherical halo with a dark matter density profile as 
originally derived by \citet{1996ApJ...462..563N,1997ApJ...490..493N} 
for galaxies forming in a Cold Dark Matter scenario. We 
consider also the Burkert dark matter density profile  
\citep{1995ApJ...447L..25B} since this successfully fitted the rotation  
curve of dark matter dominated dwarf galaxies \citep[e.g. ][ and references 
therein]{2007MNRAS.375..199G}. 
Both models can describe the dark matter halo density profile using two  
parameters. 
The density profile proposed by \citet{1995ApJ...447L..25B} has a   
constant-density core and is given by: 

\begin{equation} 
\rho(R)={\rho_B\over (1+{R\over R_B})\Bigl(1+({R\over R_B})^2\Bigr)} 
\end{equation}

\noindent 
A strong correlation between the two parameters $\rho_B$ and $R_B$ has 
been found by fitting the rotation curve of low mass disk galaxies
\citep{2000ApJ...537L...9S}, namely  

\begin{equation} 
M_B=4.3\times 10^7 {R_B\over {\hbox{kpc}}}^{7/3}\ M_\odot \qquad {\hbox{with }} 
M_B=1.6\rho_B R_B^3 
\end{equation} 

\begin{figure} 
\includegraphics[width=84mm]{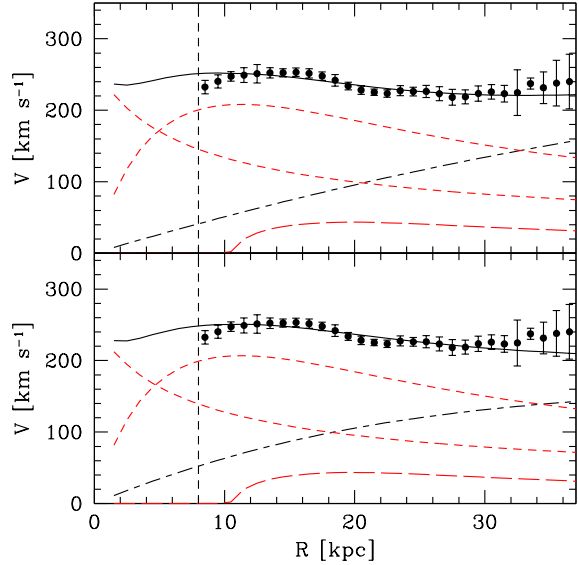} 
\caption{The M31 rotation curve (points) and the best-fitting 
mass models (solid line) using a Burkert dark halo profile with  
$h_d=5.1$~kpc, $h_b=2$~kpc and n=4. Also shown are  
the dark halo contribution (dot-dashed line), the stellar disk  
and bulge (short-dashed line)  and the gas contribution  
(long-dashed line). In the top panel, we show the best fit mass
model ($\chi^2=0.81$) with $(M/L)_b=4.5$~M$_\odot$/L$_\odot$,   
$(M/L)_d=8.0$~M$_\odot$/L$_\odot$ and $R_B=77$~kpc. The case shown in 
the bottom panel refers to a fixed, lower value of the core radius,
namely  $R_B=28$~kpc. For this case the best fitting values of
the mass-to-light ratios are $(M/L)_b=4.9$~M$_\odot$/L$_\odot$
and $(M/L)_d=7.4$~M$_\odot$/L$_\odot$ ($\chi^2=1.17$). } 
\label{burk} 
\end{figure} 

\begin{figure} 
\includegraphics[width=84mm]{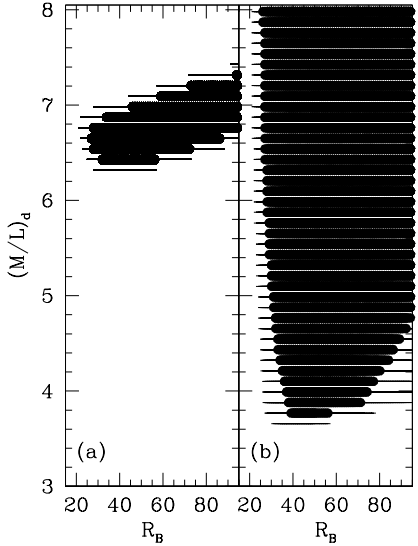} 
\caption{The 68~$\%$ (darker regions) and 95.4~$\%$ (lighter regions)  
confidence areas in the $(M/L)_d$--$R_B$ plane. Clearly the confidence
areas extend even beyond the largest $R_B$ value shown.
In panel $(a)$ $(M/L)_d=(M/L)_b$, 
while in $(b)$ $(M/L)_d$ and $(M/L)_b$ are two independent variables. 
} 
\label{prob_b} 
\end{figure}

Using this relation, 
we show in Figure~\ref{burk} the best fitting mass model when a 
Burkert model for the dark halo is considered. 
The values of the 
free parameters are: $R_B=77$~kpc, $(M/L)_d=8.0$~M$_\odot$/L$_\odot$
and $h_d=5.1$~kpc. The reduced $\chi^2$ value is close to unity, 
$\chi^2=0.81$, meaning the fit is generally good.  
The best fit $\chi^2$ does not depend much on the assumed
bulge parameters. Even considering $(M/L)_d=(M/L)_b$ the halo model
provides an excellent fit to the data. We define the virial
mass, $M_{B}^{vir}$, in the case of a Burkert halo as the mass enclosed
in the sphere which has the average dark matter density equal to 
98~$\rho_{cr}$
where the critical density for closure is $\rho_{cr}=3 H_0^2/8\pi G$
with $H_0=71$~km~s$^{-1}$. The virial mass of the dark halo for
best fitting mass model is uncomfortably high: 
$M_{B}^{vir}=1.2\ 10^{13}$~M$_\odot$.
A Burkert halo which fits M31 is indistinguishable from a  
constant density dark matter model since 
even at the outermost fitted radius, the dark matter  
density does not decline yet as an $R^{-2}$ or $R^{-3}$ power law. 
If we consider  core radii smaller than the best fitting value, 
comparable to the extent of the HI disk, Burkert 
halo models still provides acceptable fits to the data
and implies lower virial masses. 
In Figure~\ref{prob_b} we show the 68~$\%$ and 95.4~$\%$   
confidence areas in the $(M/L)_d$--$R_B$ plane.
For $22\le R_B \le 37$~kpc we have virial masses 
$8.3\ 10^{11}\le M_{B}^{vir}\le 2.5\ 10^{12}$~M$_\odot$ and 
$\chi^2 $ values in the 95.4~$\%$ confidence area
($\chi^2$ increases as $R_B$ decreases). In the bottom panel
of Figure~\ref{burk} we show, as an example, a similar mass model to
that shown in the top panel, except for the value of the core radius 
which we set equal to 28~kpc. With such low value of $R_B$
the Burkert halo model has a virial mass of $M_B^{vir}=1.4\ 10^{12}$~M$_\odot$ 
and provides still an acceptable fit to the data ($\chi^2=1.17$).

The NFW density profile is: 

\begin{equation} 
\rho(R)={\rho_{NFW} \over {R\over R_{NFW}}\Bigl(1+{R\over R_{NFW}}\Bigr)^2} 
\end{equation} 

\noindent 
Numerical simulations of galaxy formation find a  
correlation between $\rho_{NFW}$ and $R_{NFW}$ which depends on the  
cosmological model \citep[e.g. ][]{1997ApJ...490..493N,2001ApJ...559..516A, 
2001ApJ...554..114E,2001MNRAS.321..559B}. 
Often this correlation is expressed using the concentration parameter  
$C\equiv R_{\Delta}/R_{NFW}$ and $M_{\Delta}$ or $V_{\Delta}$.  
$R_{\Delta}$ is the radius of a sphere containing a mean density $\Delta$  
times the cosmological critical density, $V_{\Delta}$ and $M_{\Delta}$ are the  
characteristic velocity and mass inside $R_{\Delta}$. In this paper 
we use the results of N-body simulations in $\Lambda$CDM,
cosmology to define the relation 
between the concentration and the virial mass of dark matter halos. 
We adopt a flat $\Lambda$CDM cosmology with parameters from the  
$WMAP$ results \citep{2003ApJS..148..175S}, matter density $\Omega_M=0.27$, 
baryon density $\Omega_b=0.044$, a normalized Hubble constant  
$h\equiv H_0/$(100~km~s$^{-1}$~Mpc$^{-1}$)=0.71. In this framework, we use 
the results of the numerical simulation of \citet{2007MNRAS.378...55M},
which are in agreement with the results of \citet{2007MNRAS.381.1450N}, 
and give the following  relation between the concentration and the halo 
mass for relaxed halos: 

\begin{equation} 
C= 7.5 (M_{\Delta}/10^{14} h^{-1} M_\odot)^{-0.098} 
\end{equation} 

\noindent 
where $M_{\Delta}$ is the halo mass for a density contrast $\Delta=98$. 
Generally the dispersion in the
concentration decreases monotonically as a function of halo mass.
The scatter around log$C$ given by the above relation, 
is about $\pm 0.13$ at the expected virial mass of Andromeda 
(i.e. $10^{12}$~M$_\odot$), and a similar value is found for the
average total scatter over the mass range considered.

We now fit the M31 rotation curve using two free parameters: the 
halo concentration C and $(M/L)_d$, the disk mass-to-light ratio. 
We shall consider first no dispersion around the relation given by the 
above equation and a bulge mass-to-light ratio equal to that of the disk. 
The 1-$\sigma$, 2-$\sigma$ and 3-$\sigma$ ranges are determined  using the  
reduced $\chi^2$ and assuming Gaussian statistics. They are computed by  
determining in the free parameter space the projection ranges, along each  
axis, of the hypersurfaces corresponding to the 68.3$\%$ and 95.4$\%$ 
confidence levels. 

\begin{figure} 
\includegraphics[width=84mm]{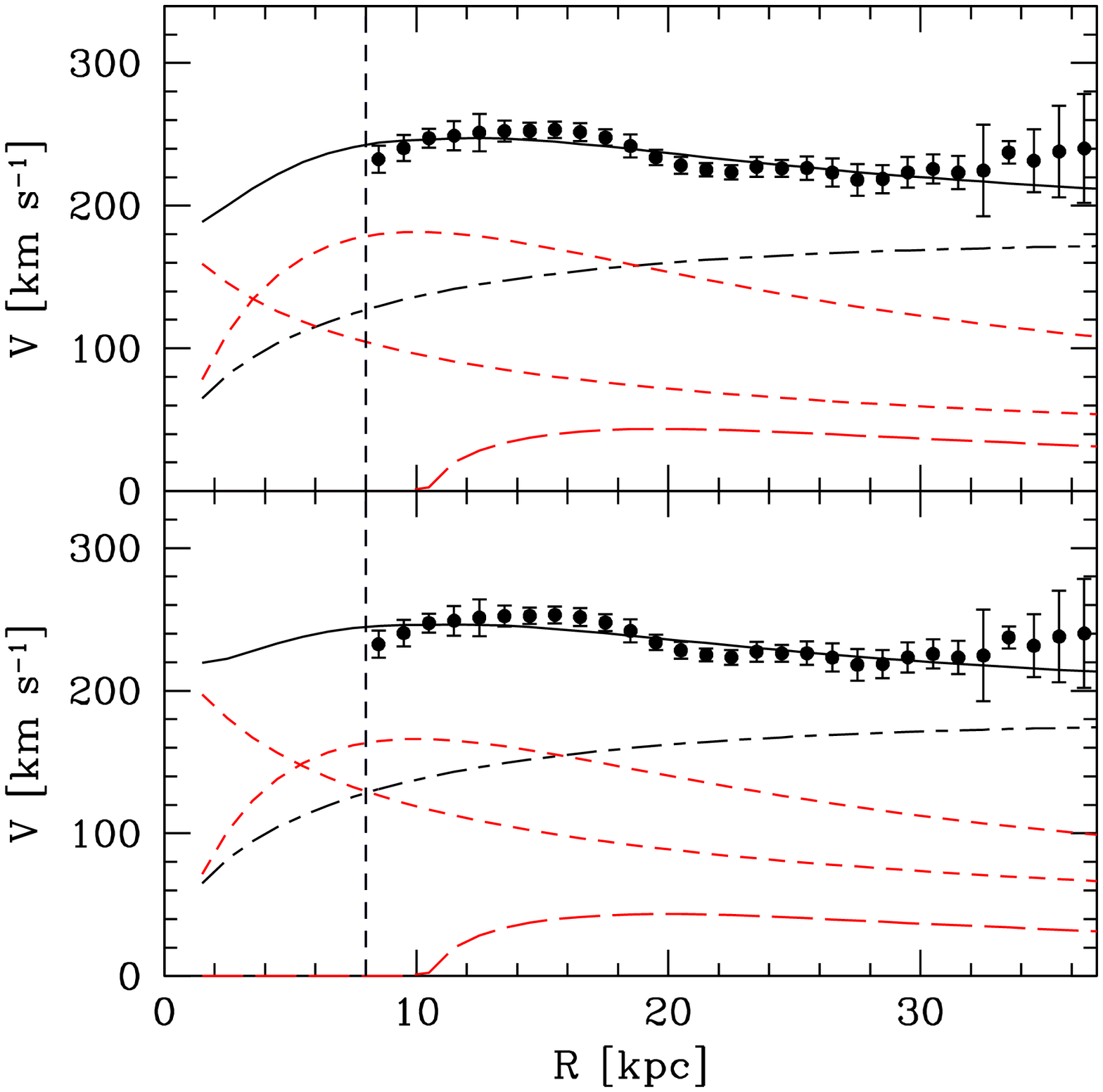} 
\caption{The M31 rotation curve (points) and the best-fitting 
mass model (solid line) using the NFW dark halo profile with $C=12$ 
in the frame of $\Lambda CDM$. Also shown are  
the dark halo contribution (dot-dashed line), the stellar disk  
and bulge (short-dashed line)  and the gas contribution  
(long-dashed line). The bottom panel refers to the 
case ($(M/L)_d=(M/L)_b=4.2$~M$_\odot$/L$_\odot$). The top 
panel refers to the best fit when the mass-to-light ratio 
of the disk and the bulge are two independent variables. For 
the best fit $(M/L)_d=5.0$ and an $(M/L)_b=2.7$~M$_\odot$/L$_\odot$.
 These combination of  M/L
ratios is not realistic since we don't expect the bulge to have a lower
M/L ratio than the disk. But higher values of the bulge M/L ratio increases
only slightly the $\chi^2$ value and require similar dark matter distribution. 
Both fits refer to a bulge model with $h_b=2.0$~kpc and n=4.
} 
\label{lcdm} 
\end{figure}

The best fitting mass model is obtained for a disk scalelength $h_d=4.5$~kpc. 
The bulge parameters are not of much relevance for the goodness of the fit in
the region of interest to this paper.
The stellar mass-to-light ratio for the best fit is  
$(M/L)_d=4.2$~M$_\odot$/L$_\odot$, if we assume that it does not
vary between the bulge and the disk component, and the value of the reduced  
$\chi^2$ is 1.12. The total dark halo mass is  
$M_\Delta=10^{12}$~M$_\odot$ (corresponding to $C=11.9$). 
If we allow variations
between the disk and the bulge mass-to-light ratio, the best fit mass model gives
$(M/L)_d=5.0$~M$_\odot$/L$_\odot$ and $(M/L)_b=2.7$~M$_\odot$/L$_\odot$
and a minimum $\chi^2$ value of 1.08.  These combination of  M/L
ratios is not realistic since we don't expect the bulge to have a lower
M/L ratio than the disk. But higher values of the bulge M/L ratio increases
only slightly the $\chi^2$ value.
In Figure~\ref{lcdm} we show the modelled rotation curve according to 
the best fitting mass model when $h_b=2.0$~kpc and n=4. The
dark halo mass is $M_\Delta=1.2\times10^{12}$~M$_\odot$ corresponding to $C=12$.
As we increase the disk scalelength the $\chi^2$ increases slightly, 
the dark halo mass 
decreases and the stellar mass-to-light ratio increases  
(for example for a change in the disk 
scalelength from 4.5 to 6.1~kpc the minimum $\chi^2$ changes from 
1.12 to 1.34, $M_\Delta$ decreases from $1.2\times10^{12}$ to  
$7.5\times 10^{11}$~M$_\odot$ and $(M/L)_{b,d}=5.7$~M$_\odot$/L$_\odot$).

\begin{figure} 
\includegraphics[width=84mm]{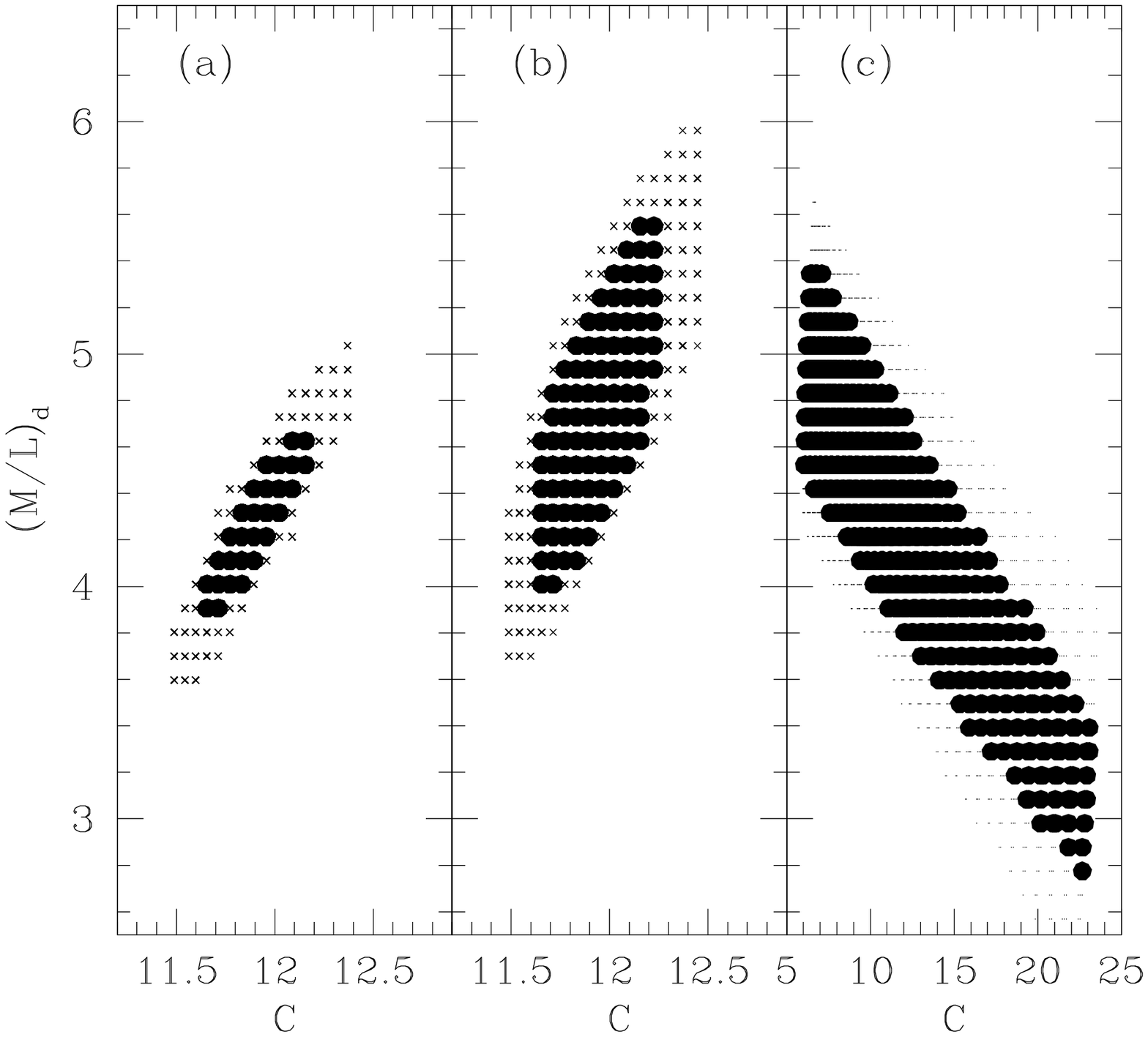} 
\caption{The 68~$\%$ (darker regions) and 95.4~$\%$ (lighter regions)  
confidence areas in the $(M/L)_d$--C plane. Panel $(a)$ and $(b)$ refers  
to the case where the average C-M$_\Delta$ relation is considered  
with no dispersions around it. In panel $(a)$ $(M/L)_d=(M/L)_b$, 
while in $(b)$ $(M/L)_d$ and $(M/L)_b$ are two independent variables. 
In panel $(c)$ the confidence areas are for the $(M/L)_d=(M/L)_b$ case 
when 2-$\sigma=0.26$ around the average logC-logM$_\Delta$ relation  
is considered.} 
\label{prob} 
\end{figure}

We have computed the confidence area in the $(M/L)_d-C$ plane and shown 
them in Figure~\ref{prob}.  When the mass-to-light ratio 
of the disk and the bulge are two independent variables the best fitting
$(M/L)_b$ value is unrealistically low but the value of the concentration
parameters does not change. Figure~\ref{prob} shows also the wider 
confidence areas obtained  when twice the dispersion of $\pm 0.13$, as estimated 
from numerical simulations, is considered around log$C$ in the  
C-M$_\Delta$ relation. 

\begin{table*}
\caption{Parameters of the best fitting mass model.}
\label{massmod}
\begin{minipage}{\textwidth}
\begin{center}
\begin{tabular}{lccccccc}
\hline \hline
     MODEL &  n$_b$ &  h$_b$  & h$_d$ &  (M/L)$_b$ &  (M/L)$_d$  &  $P^+$  & $\chi^2$ \\
     &  &  kpc  & kpc &  (M/L)$_\odot$ &  (M/L)$_\odot$  &  & \\
\hline \hline

No-DM  & 4 & 0.7     & 6.1 & 8.0 & 8.0 & ... & 6.10\\
No-DM  & 4 & 2.0     & 6.1 & 8.0 & 8.0 & ... & 6.15\\
No-DM  & 1.6 & 0.7   & 6.1 & 8.0 & 8.0 & ... & 6.17\\
No-DM  & 1.6 & 2.0   & 6.1 & 8.0 & 8.0 & ... & 6.02\\
MOND   & 4 & 0.7     & 4.5 & 5.2 & 2.5 & ... & 1.32\\    
MOND   & 4 & 2.0     & 4.5 & 5.6 & 2.5 & ... & 1.41\\ 
MOND   & 1.6 & 0.7   & 4.5 & 5.1 & 2.5 & ... & 1.31\\ 
MOND   & 1.6 & 2.0   & 4.5 & 5.2 & 2.5 & ... & 1.30\\ 
Burkert  & 4 & 0.7   & 5.3 & 4.6 & 8.0 & $R_B$=83 kpc & 0.84\\ 
Burkert  & 4 & 2.0   & 5.1 & 4.5 & 8.0 & $R_B$=77 kpc & 0.81\\ 
Burkert  & 1.6 & 0.7 & 5.3 & 4.5 & 8.0 & $R_B$=74 kpc  & 0.84\\ 
Burkert  & 1.6 & 2.0 & 5.3 & 4.5 & 8.0 & $R_B$=83 kpc  & 0.83\\ 
$\Lambda$CDM  & 4 & 0.7    & 4.5 & 2.6 & 4.9 & C=11.9 & 1.09\\
$\Lambda$CDM  & 4 & 2.0    & 4.5 & 2.7 & 5.0 & C=12.0 & 1.08\\
$\Lambda$CDM  & 1.6 & 0.7  & 4.5 & 2.6 & 4.9 & C=11.9 & 1.10\\
$\Lambda$CDM  & 1.6 & 2.0  & 4.5 & 2.6 & 4.9 & C=11.9 & 1.08\\  
\hline
MOND   & 4 & 0.7     & 4.5 & 3.3 & 3.3 & ... & 1.37\\    
MOND   & 4 & 2.0     & 4.5 & 3.3 & 3.3 & ... & 1.48\\ 
MOND   & 1.6 & 0.7   & 4.5 & 3.3 & 3.3 & ... & 1.37\\
MOND   & 1.6 & 2.0   & 4.5 & 3.3 & 3.3 & ... & 1.37\\
Burkert  & 4 & 0.7   & 5.5 & 7.0 & 7.0 & $R_B$=74 kpc & 0.90\\ 
Burkert  & 4 & 2.0   & 5.3 & 7.0 & 7.0 & $R_B$=78 kpc & 0.85\\ 
Burkert  & 1.6 & 0.7 & 5.5 & 6.9 & 6.9 & $R_B$=66 kpc & 0.92\\ 
Burkert  & 1.6 & 2.0 & 5.5 & 7.0 & 7.0 & $R_B$=76 kpc & 0.89\\ 
$\Lambda$CDM  & 4 & 0.7 &  4.5  & 4.1 & 4.1 & C=11.8 & 1.14\\
$\Lambda$CDM  & 4 & 2.0 &  4.5  & 4.2 & 4.2 & C=11.9 & 1.12\\
$\Lambda$CDM  & 1.6 & 0.7 & 4.5 & 4.0 & 4.0 & C=11.8 & 1.15\\
$\Lambda$CDM  & 1.6 & 2.0 & 4.5 & 4.1 & 4.1 & C=11.8 & 1.12\\

\hline \hline
\end{tabular}
\end{center}
\end{minipage}
\end{table*}

Table~\ref{massmod} summarizes the main results of the mass models
considered for fitting the rotation curve of M31. In column (1) we give 
the short name of the mass model: No-DM is simply a Newtonian
dynamic fit with no dark matter. For MOND we use the no dark matter and
the ``simple'' interpolation function.
The scalelengths of the bulge and of the disc are labeled with the symbol
$h_b$ and $h_d$ respectively. We consider h$_d$ as a free parameter
in the interval 4.5-6.1~kpc. The best fitting value of a possible additional  
free parameter of each model, P$^+$ is given in column (7). For the No-DM and
MOND models there is no additional free parameter. For the Burkert halo model 
P$^+$ is the core radius in kpc, for the $\Lambda$CDM mass model P$^+$ is
the concentration parameter C. For the bulge and disk mass-to-light
ratio, the range of possible values considered are: 2.5$\le$(M/L)$_b\le$ 8.0
in units of (M/L)$_\odot$.
In the first half of the Table, we consider both (M/L)$_b$ and (M/L)$_d$
as free parameters while in the second part we use (M/L)$_b$=(M/L)$_d$. 
Since, for
the simple Newtonian dynamic fit with no dark matter, the bulge and disk
mass-to-light ratios come out equal to the highest allowed value, 
the parameters are listed only in the first part of the Table. 

We can see clearly in Table~\ref{massmod} that
the goodness of the fit does not
depend on the bulge mass distribution, as expected, since we are only
fitting the rotation curve for $R>8$~kpc. It also shows that only a
pure Newtonian mass model with no dark matter halo fails to fit the data.
Both MOND, a Burkert halo model with a large constant density
core, and the NFW profile in the framework 
of $\Lambda CDM$ provide good mass model fits to the rotation curve 
of Andromeda derived in this paper. In the 
next subsection we shall use data from other sources to test the 
rotation curve at larger radii than those provided by the HI dataset.

\subsection{Baryonic and total mass of Andromeda}

\begin{figure} 
\includegraphics[width=96mm]{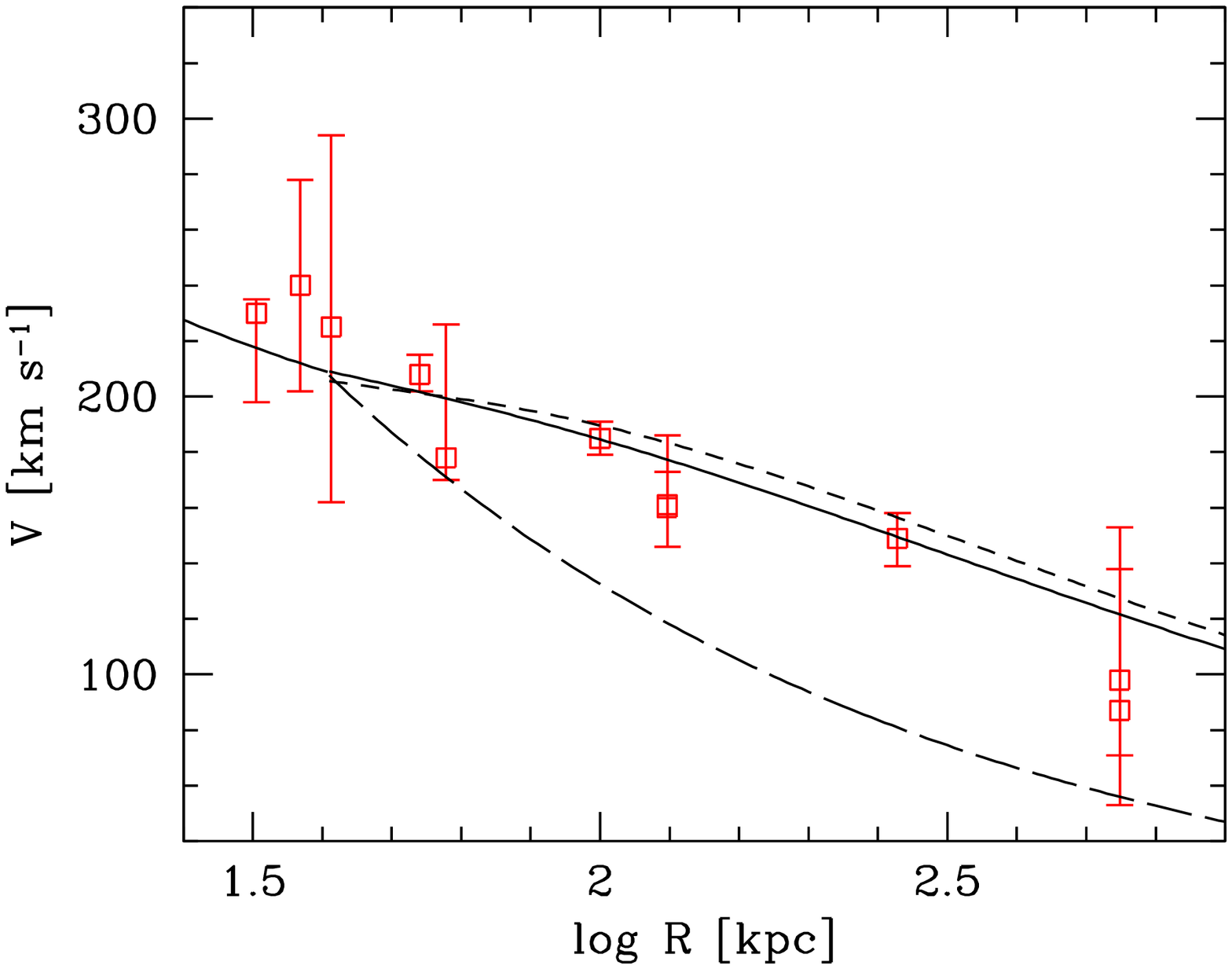} 
\caption{ Rotational velocities predicted 
at large distances from the center of Andromeda according to several 
datasets analyzed by previous papers (see Table 3 for details). 
The continuous line shows the rotational velocities predicted by an 
extrapolation at large radii of the best fitting mass model with a 
NFW dark halo profile ($C=12$). The short dashed line shows the predicted 
velocities of a Burkert halo model with $R_B=28$~kpc. The  
long dashed line is for a constant halo dark matter density profile which 
is truncated outside 37~kpc and gives a Keplerian fall off of the velocity 
at larger distances.
} 
\label{large} 
\end{figure}

We consider now the mass estimate of Andromeda from different sources 
at galactocentric radii from 30 to 560~kpc. We derive a rotation 
curve by computing the expected circular rotational velocity 
given the mass estimate within a radius $R$. 
We consider data from planetary nebulae, globular clusters, stellar streams  
and Andromeda satellites  to constrain the Andromeda total mass
at large radii. Since each paper considers an ensemble of objects and 
describes in detail the method used to derive the mass, we give in Table 3 
the resulting mass estimate and 
reference the original papers where a more detailed description of the 
database and analysis can be found. If the authors have assumed a different
distance to Andromeda than what we use in this paper, we
have scaled their radius and mass estimate accordingly.
In Figure~\ref{large} we plot the rotational velocities  
at large distances from the center of Andromeda derived from the
mass estimates given in Table 3.

We plot also the velocities at large radii predicted by the $\Lambda CDM$ mass models 
which best fits our rotation curve i.e.
a NFW dark halo profile with concentration $C=12$ (continuous line). A Burkert halo 
model with a very large core radius, such as that required by the best fitting mass
model to the 21-cm rotation curve data, predicts higher velocities than observed at
distances shown in Figure~\ref{large}. But a Burkert halo profile with a core
radius $R_B=28$~kpc,
compatible with the 21-cm rotation curve data, is also compatible with data at  
large galactocentric distances
(short dashed line in Figure~\ref{large}). For both halo
models we have considered the values of the free parameters  $h_d$, $(M/L)_d$ and 
$(M/L)_b$, which best fit 21-cm rotation curve.  
Given the fact that the $\Lambda CDM$
and Burkert halo
were constrained using only HI data between 8 and 37~kpc
it is quite remarkable how well the two halo models  fit the data
out to about 560~kpc.  We did not test MOND predictions in this framework 
because Andromeda masses in the original paper have been derived using
Newtonian dynamic. 

Since the best fit Burkert halo model was essentially 
a constant density profile in the region traced by the 21-cm rotation curve,
we show also the predicted velocities of this model 
in case the dark halo is truncated beyond 37~kpc.
A simple keplerian fall off of the observed velocities outside the
region covered by our 21-cm dataset however fails to fit the data at larger
galactocentric radii.  Only the outermost 3 data points are compatible
with a keplerian fall-off regime. For the $\Lambda CDM$ mass model, the virial
radius corresponding to a concentration $C=12$ is 270~kpc, hence the predicted
rotational velocities of the outermost data points in Figure~\ref{large} are in 
the keplerian regime.

The best fitting $\Lambda CDM$ mass model implies a stellar (disk+bulge) mass of 
$1.3\times 10^{11}$~M$_\odot$ and a dark matter halo mass of 
$1.2\times10^{12}$~M$_\odot$. Adding to the stellar mass the cold gas 
mass (neutral and molecular hydrogen plus helium) of   
$7.7\times 10^9$~M$_\odot$ we estimate a total baryonic mass of Andromeda 
of $1.4\times 10^{11}$~M$_\odot$. This adds
to the dark matter halo mass, giving

\begin{equation} 
M_{tot}=1.3\times 10^{12}\ {\hbox{M}}_\odot 
\end{equation} 

as our best estimate of the total mass of the Andromeda galaxy. The
associated baryonic fraction is 0.12, very similar to the cosmic
inferred value of 0.14.

\begin{table*}
\caption{Andromeda mass estimated at large galactocentric radii.}
\label{largedat}
\begin{minipage}{\textwidth}
\begin{center}
\begin{tabular}{lcccc}
\hline \hline
     Reference &  R &  Mass  & V$_c$  &  Objects \\
              &  [kpc] & [10$^{10}$~M$_\odot$] & km~s$^{-1}$ & \\
\hline \hline

\citet{2000MNRAS.316..929E} & 32 & 39$^{+2}_{-10}$ & 230$^{+5}_{-32}$ & 17 Globular Clusters
\footnote{Best mass estimate using GCs; the lower mass extreme is the value inferred using 9 PNe, 
the upper mass extreme is from data on 11 Satellites} \\
This paper                  & 37 & 49$^{+12}_{-14}$ & 240$^{+38}_{-38}$ & 21-cm data \\
\citet{2000MNRAS.316..929E} & 41 & 48$^{+34}_{-23}$ & 225$^{+69}_{-63}$  & 17 Globular Clusters
\footnote{Uncertanties derived from ``Error analysis'' section of original paper} \\
\citet{2008ApJ...674..886L} & 55 & 55$^{+4}_{-3}$   & 208$^{+7}_{-6}$    & 504 Globular Clusters \\
Galleti et al. (2006) & 60 & 44$^{+26}_{-4}$  & 178$^{+48}_{-8}$  & 349 Globular Clusters
\footnote{Uncertanties include uncertanties on the isotropy degree of the velocity 
distribution and the mass estimate using only the 14 most distant GCs} \\
\citet{2000ApJ...537L..91C} & 100 & 79$^{+5}_{-5}$ & 185$^{+6}_{-6}$ & 12 Satellites
\footnote{Mass estimate assuming isotropic orbits; the mass range is larger, 
79$^{+174}_{-46}$, if models with only radial or circular orbits are included} \\
\citet{2004MNRAS.351..117I} & 125 & 75$^{+25}_{-13}$  & 161$^{+25}_{-15}$  & Stellar Stream \\
\citet{2006MNRAS.366.1012F} & 125 & 74$^{+12}_{-12}$  & 160$^{+13}_{-14}$  & Stellar Stream \\
\citet{1999AJ....118..337C} & 268 & 137$^{+18}_{-18}$ & 149$^{+9}_{-10}$   & 7 Satellites \\
\citet{2000MNRAS.316..929E} & 560 & 125$^{+180}_{-60}$ & 98$^{+55}_{-27}$  & 11 Satellites
\footnote{Uncertanties derived from ``Error analysis'' section of original paper } \\
\citet{2000ApJ...540L...9E} & 560 & 99$^{+146}_{-63}$  & 87$^{+51}_{-34}$  & 16 Satellites
\footnote{Mass estimate assuming Osipkov-Merrit distribution function; the given mass range
includes also the mass estimated assuming a constant orbital anisotropy.} \\
\hline \hline
\end{tabular}
\end{center}
\end{minipage}
\end{table*}


\section{Summary} 

   We determine the rotation curve of the M31 disk from 8 to 37~kpc using a 
   tilted ring model fit to the velocity field mapped in the
   full-disk, 21-cm  
   imaging survey of \citet{2009ApJ...695..937B}. The orientation of 
   the rings have been determined using three different
   techniques which give rather similar results. For our most accurate modelling 
   method (P1), we use 11 equally spaced 
   free rings, which cover galactocentric distances between 9 and 36~kpc, whose 
   parameters are varied independently. Each free ring has 6 degrees of
   freedom, since we allowed the systemic velocity and center position of
   each ring to vary (in addition to the circular rotation, inclination
   and position angle).  This implies a total of 66 degrees of freedom in our
   model. Between two consecutive free rings, parameters are 
   determined by a linear interpolation. We find that the disk of M31 warps  
   from 25~kpc outwards by lowering its position angle and becoming more inclined 
   with respect to our line of sight. The disk reaches an inclination of
    86$^\circ$ at 35~kpc. The geometry of the outermost
   two rings has somewhat larger uncertainties, but the tilted ring model
   which gives the best fit to the data also produces consistent
   rotation curves in the two separate halves of the galaxy. Furthermore,   
   these rotation curves  do not depend on the value of limiting 
   angle around the major axis chosen for selecting the data. We find 
   -306~km~s$^{-1}$ as the average value of the systemic velocity of the
   gaseous disk of M31. The rotation curve of M31 is consistent with
   being flat beyond 20~kpc and we carry on a dynamical analysis to
   determine the baryonic and non-baryonic mass distribution of the 
   nearest spiral galaxy.

   The M31 rotation curve cannot be reproduced using Newtonian dynamics
   and only the stellar and gaseous mass components. Without a dark matter halo
   however,  MOND provide a good fit to the galaxy gravitational 
   potential in the region considered. We test the density 
   profile and mass predictions of hierarchical clustering and structure 
   formation in a $\Lambda$CDM cosmology, together with a dark halo 
   model having a constant density core. Both models are able 
   to reproduce the rotation curve of M31 to a high level of accuracy.
   The constant density core model which fits M31 has a core radius
   comparable to the size of the disk of M31 and therefore is in practice a
   constant dark matter density model. 

   Using the relation between the concentration parameter C and the dark halo
   mass M$_\Delta$ as for a NFW density profile in a flat $\Lambda$CDM cosmology, we
   find a best fit halo concentration parameter C=12 which implies a dark 
   matter halo mass M$_\Delta$=1.2$\times10^{12}$~M$_\odot$. 
   If we assume that the stellar disk and the bulge have the same mass-to-light
   ratio we find $(M/L)_d=(M/L)_b=4.2\pm 0.4 $~M$_\odot$/L$_\odot$. If the 
   mass-to-light ratio of the disk and the bulge are two independent variables then 
   the best fit gives a slightly higher value for the disk, $(M/L)_d=5.0^{+0.6}_{-1.0}$.
   We are unable to constrain the bulge mass-to-light ratio 
   since we discarded the innermost rotation curve in our fit
   (derived without considering elliptical orbits).  
   A wider range of C and $(M/L)_d$ values are
   found when a dispersion is considered  around the average 
   log C -- log M$_\Delta$ relation, as suggested by numerical simulations of
   structure formation in a $\Lambda$CDM cosmology.

   An interesting result is that 
   the extrapolation of a constant core dark halo model, 
   as well as of the best fit $\Lambda CDM$ dark halo model, beyond the region 
   traced by the rotation curve are in good agreement
   with the Andromeda mass traced by other dynamical indicators 
   (globular clusters, streams, satellites) at larger radii, out to 560~kpc.
   The constant-core best fitting halo model has a very large core radius (77~kpc) and a
   high virial mass, not consistent with the data at large galactocentric radii. However
   models with a somewhat smaller core radius, provide acceptable fit to the 21-cm
   rotation curve and to data at larger galactocentric radii.
   The total estimated mass of M31 from our 
   mass model fit to the 21-cm rotation curve in the framework of $\Lambda CDM$ cosmology
   is 1.3$^{+0.3}_{-0.3}\times 10^{12}$~M$_\odot$, with a 0.12 baryonic fraction.
   This is similar to the cosmic inferred baryonic fraction of 0.14 and implies a formation
   redshift $z_f=1.2$ for the Andromeda galaxy. 


\begin{acknowledgements} 
RW acknowledges support from Research Corporation for the Advancement of 
Science. We acknowledge the anonymous referee for his/her criticism to
an earlier version of the paper.
\end{acknowledgements} 



\begin{thebibliography}{85}
\expandafter\ifx\csname natexlab\endcsname\relax\def\natexlab#1{#1}\fi

\bibitem[{{Athanassoula} \& {Beaton}(2006)}]{2006MNRAS.370.1499A}
{Athanassoula}, E. \& {Beaton}, R.~L. 2006, \mnras, 370, 1499

\bibitem[{{Avila-Reese} {et~al.}(2001){Avila-Reese}, {Col{\'{\i}}n},
  {Valenzuela}, {D'Onghia}, \& {Firmani}}]{2001ApJ...559..516A}
{Avila-Reese}, V., {Col{\'{\i}}n}, P., {Valenzuela}, O., {D'Onghia}, E., \&
  {Firmani}, C. 2001, \apj, 559, 516

\bibitem[{{Babcock}(1939)}]{1939LicOB..19...41B}
{Babcock}, H.~W. 1939, Lick Observatory Bulletin, 19, 41

\bibitem[{{Bajaja} \& {Shane}(1982)}]{1982A&AS...49..745B}
{Bajaja}, E. \& {Shane}, W.~W. 1982, \aaps, 49, 745

\bibitem[{{Barmby} {et~al.}(2006){Barmby}, {Ashby}, {Bianchi}, {Engelbracht},
  {Gehrz}, {Gordon}, {Hinz}, {Huchra}, {Humphreys}, {Pahre},
  {P{\'e}rez-Gonz{\'a}lez}, {Polomski}, {Rieke}, {Thilker}, {Willner}, \&
  {Woodward}}]{2006ApJ...650L..45B}
{Barmby}, P., {Ashby}, M.~L.~N., {Bianchi}, L., {et~al.} 2006, \apjl, 650, L45

\bibitem[{{Battaner} {et~al.}(1986){Battaner}, {Beckman}, {Mediavilla},
  {Prieto}, {Sanchez Magro}, {Munoz Tunon}, \& {Sanchez
  Saavedra}}]{1986A&A...161...70B}
{Battaner}, E., {Beckman}, J.~E., {Mediavilla}, E., {et~al.} 1986, \aap, 161,
  70

\bibitem[{{Beaton} {et~al.}(2007){Beaton}, {Majewski}, {Guhathakurta},
  {Skrutskie}, {Cutri}, {Good}, {Patterson}, {Athanassoula}, \&
  {Bureau}}]{2007ApJ...658L..91B}
{Beaton}, R.~L., {Majewski}, S.~R., {Guhathakurta}, P., {et~al.} 2007, \apjl,
  658, L91

\bibitem[{{Begeman}(1987)}]{1987PhDT.......199B}
{Begeman}, K.~G. 1987, PhD thesis, , Kapteyn Institute, (1987)

\bibitem[{{Begeman}(1989)}]{1989A&A...223...47B}
{Begeman}, K.~G. 1989, \aap, 223, 47

\bibitem[{{Bell} \& {de Jong}(2001)}]{2001ApJ...550..212B}
{Bell}, E.~F. \& {de Jong}, R.~S. 2001, \apj, 550, 212

\bibitem[{{Berman}(2001)}]{2001A&A...371..476B}
{Berman}, S. 2001, \aap, 371, 476

\bibitem[{{Berman} \& {Loinard}(2002)}]{2002MNRAS.336..477B}
{Berman}, S. \& {Loinard}, L. 2002, \mnras, 336, 477

\bibitem[{{Braun}(1991)}]{1991ApJ...372...54B}
{Braun}, R. 1991, \apj, 372, 54

\bibitem[{{Braun} {et~al.}(2009){Braun}, {Thilker}, {Walterbos}, \&
  {Corbelli}}]{2009ApJ...695..937B}
{Braun}, R., {Thilker}, D.~A., {Walterbos}, R.~A.~M., \& {Corbelli}, E. 2009,
  \apj, 695, 937

\bibitem[{{Brinks} \& {Burton}(1984)}]{1984A&A...141..195B}
{Brinks}, E. \& {Burton}, W.~B. 1984, \aap, 141, 195

\bibitem[{{Bullock} {et~al.}(2001){Bullock}, {Kolatt}, {Sigad}, {Somerville},
  {Kravtsov}, {Klypin}, {Primack}, \& {Dekel}}]{2001MNRAS.321..559B}
{Bullock}, J.~S., {Kolatt}, T.~S., {Sigad}, Y., {et~al.} 2001, \mnras, 321, 559

\bibitem[{{Burkert}(1995)}]{1995ApJ...447L..25B}
{Burkert}, A. 1995, \apjl, 447, L25+

\bibitem[{{Carignan} {et~al.}(2006){Carignan}, {Chemin}, {Huchtmeier}, \&
  {Lockman}}]{2006ApJ...641L.109C}
{Carignan}, C., {Chemin}, L., {Huchtmeier}, W.~K., \& {Lockman}, F.~J. 2006,
  \apjl, 641, L109

\bibitem[{{Casertano}(1983)}]{1983MNRAS.203..735C}
{Casertano}, S. 1983, \mnras, 203, 735

\bibitem[{{Chapman} {et~al.}(2008){Chapman}, {Ibata}, {Irwin}, {Koch},
  {Letarte}, {Martin}, {Collins}, {Lewis}, {McConnachie}, {Pe{\~n}arrubia},
  {Rich}, {Trethewey}, {Ferguson}, {Huxor}, \& {Tanvir}}]{2008MNRAS.390.1437C}
{Chapman}, S.~C., {Ibata}, R., {Irwin}, M., {et~al.} 2008, \mnras, 390, 1437

\bibitem[{{Chemin} {et~al.}(2009){Chemin}, {Carignan}, \&
  {Foster}}]{2009ApJ...705.1395C}
{Chemin}, L., {Carignan}, C., \& {Foster}, T. 2009, \apj, 705, 1395

\bibitem[{{Corbelli}(2003)}]{2003MNRAS.342..199C}
{Corbelli}, E. 2003, \mnras, 342, 199

\bibitem[{{Corbelli} \& {Salpeter}(1993)}]{1993ApJ...419..104C}
{Corbelli}, E. \& {Salpeter}, E.~E. 1993, \apj, 419, 104

\bibitem[{{Corbelli} \& {Salucci}(2007)}]{2007MNRAS.374.1051C}
{Corbelli}, E. \& {Salucci}, P. 2007, \mnras, 374, 1051

\bibitem[{{Corbelli} \& {Schneider}(1997)}]{1997ApJ...479..244C}
{Corbelli}, E. \& {Schneider}, S.~E. 1997, \apj, 479, 244

\bibitem[{{Corbelli} \& {Walterbos}(2007)}]{2007ApJ...669..315C}
{Corbelli}, E. \& {Walterbos}, R.~A.~M. 2007, \apj, 669, 315

\bibitem[{{C{\^o}t{\'e}} {et~al.}(2000){C{\^o}t{\'e}}, {Mateo}, {Sargent}, \&
  {Olszewski}}]{2000ApJ...537L..91C}
{C{\^o}t{\'e}}, P., {Mateo}, M., {Sargent}, W.~L.~W., \& {Olszewski}, E.~W.
  2000, \apjl, 537, L91

\bibitem[{{Courteau} \& {van den Bergh}(1999)}]{1999AJ....118..337C}
{Courteau}, S. \& {van den Bergh}, S. 1999, \aj, 118, 337

\bibitem[{{Cram} {et~al.}(1980){Cram}, {Roberts}, \&
  {Whitehurst}}]{1980A&AS...40..215C}
{Cram}, T.~R., {Roberts}, M.~S., \& {Whitehurst}, R.~N. 1980, \aaps, 40, 215

\bibitem[{{de Blok} {et~al.}(2001){de Blok}, {McGaugh}, {Bosma}, \&
  {Rubin}}]{2001ApJ...552L..23D}
{de Blok}, W.~J.~G., {McGaugh}, S.~S., {Bosma}, A., \& {Rubin}, V.~C. 2001,
  \apjl, 552, L23

\bibitem[{{Eke} {et~al.}(2001){Eke}, {Navarro}, \&
  {Steinmetz}}]{2001ApJ...554..114E}
{Eke}, V.~R., {Navarro}, J.~F., \& {Steinmetz}, M. 2001, \apj, 554, 114

\bibitem[{{Emerson}(1976)}]{1976MNRAS.176..321E}
{Emerson}, D.~T. 1976, \mnras, 176, 321

\bibitem[{{Evans} \& {Wilkinson}(2000)}]{2000MNRAS.316..929E}
{Evans}, N.~W. \& {Wilkinson}, M.~I. 2000, \mnras, 316, 929

\bibitem[{{Evans} {et~al.}(2000){Evans}, {Wilkinson}, {Guhathakurta}, {Grebel},
  \& {Vogt}}]{2000ApJ...540L...9E}
{Evans}, N.~W., {Wilkinson}, M.~I., {Guhathakurta}, P., {Grebel}, E.~K., \&
  {Vogt}, S.~S. 2000, \apjl, 540, L9

\bibitem[{{Famaey} \& {Binney}(2005)}]{2005MNRAS.363..603F}
{Famaey}, B. \& {Binney}, J. 2005, \mnras, 363, 603

\bibitem[{{Fardal} {et~al.}(2006){Fardal}, {Babul}, {Geehan}, \&
  {Guhathakurta}}]{2006MNRAS.366.1012F}
{Fardal}, M.~A., {Babul}, A., {Geehan}, J.~J., \& {Guhathakurta}, P. 2006,
  \mnras, 366, 1012

\bibitem[{{Geehan} {et~al.}(2006){Geehan}, {Fardal}, {Babul}, \&
  {Guhathakurta}}]{2006MNRAS.366..996G}
{Geehan}, J.~J., {Fardal}, M.~A., {Babul}, A., \& {Guhathakurta}, P. 2006,
  \mnras, 366, 996

\bibitem[{{Gentile} {et~al.}(2005){Gentile}, {Burkert}, {Salucci}, {Klein}, \&
  {Walter}}]{2005ApJ...634L.145G}
{Gentile}, G., {Burkert}, A., {Salucci}, P., {Klein}, U., \& {Walter}, F. 2005,
  \apjl, 634, L145

\bibitem[{{Gentile} {et~al.}(2007){Gentile}, {Salucci}, {Klein}, \&
  {Granato}}]{2007MNRAS.375..199G}
{Gentile}, G., {Salucci}, P., {Klein}, U., \& {Granato}, G.~L. 2007, \mnras,
  375, 199

\bibitem[{{Gentile} {et~al.}(2004){Gentile}, {Salucci}, {Klein}, {Vergani}, \&
  {Kalberla}}]{2004MNRAS.351..903G}
{Gentile}, G., {Salucci}, P., {Klein}, U., {Vergani}, D., \& {Kalberla}, P.
  2004, \mnras, 351, 903

\bibitem[{{Gnedin} \& {Zhao}(2002)}]{2002MNRAS.333..299G}
{Gnedin}, O.~Y. \& {Zhao}, H. 2002, \mnras, 333, 299

\bibitem[{{Gordon} {et~al.}(2006){Gordon}, {Bailin}, {Engelbracht}, {Rieke},
  {Misselt}, {Latter}, {Young}, {Ashby}, {Barmby}, {Gibson}, {Hines}, {Hinz},
  {Krause}, {Levine}, {Marleau}, {Noriega-Crespo}, {Stolovy}, {Thilker}, \&
  {Werner}}]{2006ApJ...638L..87G}
{Gordon}, K.~D., {Bailin}, J., {Engelbracht}, C.~W., {et~al.} 2006, \apjl, 638,
  L87

\bibitem[{{Governato} {et~al.}(2009){Governato}, {Brook}, {Mayer}, {Brooks},
  {Rhee}, {Wadsley}, {Jonsson}, {Willman}, {Stinson}, {Quinn}, \&
  {Madau}}]{2009arXiv0911.2237G}
{Governato}, F., {Brook}, C., {Mayer}, L., {et~al.} 2009, ArXiv e-prints

\bibitem[{{Henderson}(1979)}]{1979A&A....75..311H}
{Henderson}, A.~P. 1979, \aap, 75, 311

\bibitem[{{Hiromoto} {et~al.}(1983){Hiromoto}, {Maihara}, {Oda}, \&
  {Okuda}}]{1983PASJ...35..413H}
{Hiromoto}, N., {Maihara}, T., {Oda}, N., \& {Okuda}, H. 1983, \pasj, 35, 413

\bibitem[{{Ibata} {et~al.}(2004){Ibata}, {Chapman}, {Ferguson}, {Irwin},
  {Lewis}, \& {McConnachie}}]{2004MNRAS.351..117I}
{Ibata}, R., {Chapman}, S., {Ferguson}, A.~M.~N., {et~al.} 2004, \mnras, 351,
  117

\bibitem[{{Ibata} {et~al.}(2005){Ibata}, {Chapman}, {Ferguson}, {Lewis},
  {Irwin}, \& {Tanvir}}]{2005ApJ...634..287I}
{Ibata}, R., {Chapman}, S., {Ferguson}, A.~M.~N., {et~al.} 2005, \apj, 634, 287

\bibitem[{{Irwin} {et~al.}(2001){Irwin}, {Ferguson}, {Tanvir}, {Ibata}, \&
  {Lewis}}]{2001INGN....5....3I}
{Irwin}, M.~J., {Ferguson}, A., {Tanvir}, N., {Ibata}, R., \& {Lewis}, G. 2001,
  The Newsletter of the Isaac Newton Group of Telescopes, 5, 3

\bibitem[{{Kent}(1983)}]{1983ApJ...266..562K}
{Kent}, S.~M. 1983, \apj, 266, 562

\bibitem[{{Kent}(1989{\natexlab{a}})}]{1989PASP..101..489K}
{Kent}, S.~M. 1989{\natexlab{a}}, \pasp, 101, 489

\bibitem[{{Kent}(1989{\natexlab{b}})}]{1989AJ.....97.1614K}
{Kent}, S.~M. 1989{\natexlab{b}}, \aj, 97, 1614

\bibitem[{{Klypin} {et~al.}(2002){Klypin}, {Zhao}, \&
  {Somerville}}]{2002ApJ...573..597K}
{Klypin}, A., {Zhao}, H., \& {Somerville}, R.~S. 2002, \apj, 573, 597

\bibitem[{{Lauer} {et~al.}(1993){Lauer}, {Faber}, {Groth}, {Shaya}, {Campbell},
  {Code}, {Currie}, {Baum}, {Ewald}, {Hester}, {Holtzman}, {Kristian}, {Light},
  {Ligynds}, {O'Neil}, \& {Westphal}}]{1993AJ....106.1436L}
{Lauer}, T.~R., {Faber}, S.~M., {Groth}, E.~J., {et~al.} 1993, \aj, 106, 1436

\bibitem[{{Lee} {et~al.}(2008){Lee}, {Hwang}, {Kim}, {Park}, {Geisler},
  {Sarajedini}, \& {Harris}}]{2008ApJ...674..886L}
{Lee}, M.~G., {Hwang}, H.~S., {Kim}, S.~C., {et~al.} 2008, \apj, 674, 886

\bibitem[{{Li} \& {White}(2008)}]{2008MNRAS.384.1459L}
{Li}, Y.-S. \& {White}, S.~D.~M. 2008, \mnras, 384, 1459

\bibitem[{{Lindblad}(1956)}]{1956StoAn..19....2L}
{Lindblad}, B. 1956

\bibitem[{{Loinard} {et~al.}(1995){Loinard}, {Allen}, \&
  {Lequeux}}]{1995A&A...301...68L}
{Loinard}, L., {Allen}, R.~J., \& {Lequeux}, J. 1995, \aap, 301, 68

\bibitem[{{Macci{\`o}} {et~al.}(2007){Macci{\`o}}, {Dutton}, {van den Bosch},
  {Moore}, {Potter}, \& {Stadel}}]{2007MNRAS.378...55M}
{Macci{\`o}}, A.~V., {Dutton}, A.~A., {van den Bosch}, F.~C., {et~al.} 2007,
  \mnras, 378, 55

\bibitem[{{McConnachie} {et~al.}(2005){McConnachie}, {Irwin}, {Ferguson},
  {Ibata}, {Lewis}, \& {Tanvir}}]{2005MNRAS.356..979M}
{McConnachie}, A.~W., {Irwin}, M.~J., {Ferguson}, A.~M.~N., {et~al.} 2005,
  \mnras, 356, 979

\bibitem[{{Milgrom}(1983)}]{1983ApJ...270..365M}
{Milgrom}, M. 1983, \apj, 270, 365

\bibitem[{{Mould} {et~al.}(2008){Mould}, {Barmby}, {Gordon}, {Willner},
  {Ashby}, {Gehrz}, {Humphreys}, \& {Woodward}}]{2008ApJ...687..230M}
{Mould}, J., {Barmby}, P., {Gordon}, K., {et~al.} 2008, \apj, 687, 230

\bibitem[{{Navarro} {et~al.}(1996){Navarro}, {Frenk}, \&
  {White}}]{1996ApJ...462..563N}
{Navarro}, J.~F., {Frenk}, C.~S., \& {White}, S.~D.~M. 1996, \apj, 462, 563

\bibitem[{{Navarro} {et~al.}(1997){Navarro}, {Frenk}, \&
  {White}}]{1997ApJ...490..493N}
{Navarro}, J.~F., {Frenk}, C.~S., \& {White}, S.~D.~M. 1997, \apj, 490, 493

\bibitem[{{Navarro} {et~al.}(2004){Navarro}, {Hayashi}, {Power}, {Jenkins},
  {Frenk}, {White}, {Springel}, {Stadel}, \& {Quinn}}]{2004MNRAS.349.1039N}
{Navarro}, J.~F., {Hayashi}, E., {Power}, C., {et~al.} 2004, \mnras, 349, 1039

\bibitem[{{Neto} {et~al.}(2007){Neto}, {Gao}, {Bett}, {Cole}, {Navarro},
  {Frenk}, {White}, {Springel}, \& {Jenkins}}]{2007MNRAS.381.1450N}
{Neto}, A.~F., {Gao}, L., {Bett}, P., {et~al.} 2007, \mnras, 381, 1450

\bibitem[{{Newton} \& {Emerson}(1977)}]{1977MNRAS.181..573N}
{Newton}, K. \& {Emerson}, D.~T. 1977, \mnras, 181, 573

\bibitem[{{Nieten} {et~al.}(2006){Nieten}, {Neininger}, {Gu{\'e}lin},
  {Ungerechts}, {Lucas}, {Berkhuijsen}, {Beck}, \&
  {Wielebinski}}]{2006A&A...453..459N}
{Nieten}, C., {Neininger}, N., {Gu{\'e}lin}, M., {et~al.} 2006, \aap, 453, 459

\bibitem[{{Olsen} {et~al.}(2006){Olsen}, {Blum}, {Stephens}, {Davidge},
  {Massey}, {Strom}, \& {Rigaut}}]{2006AJ....132..271O}
{Olsen}, K.~A.~G., {Blum}, R.~D., {Stephens}, A.~W., {et~al.} 2006, \aj, 132,
  271

\bibitem[{{Rhee} {et~al.}(2004){Rhee}, {Valenzuela}, {Klypin}, {Holtzman}, \&
  {Moorthy}}]{2004ApJ...617.1059R}
{Rhee}, G., {Valenzuela}, O., {Klypin}, A., {Holtzman}, J., \& {Moorthy}, B.
  2004, \apj, 617, 1059

\bibitem[{{Salucci} \& {Burkert}(2000)}]{2000ApJ...537L...9S}
{Salucci}, P. \& {Burkert}, A. 2000, \apjl, 537, L9

\bibitem[{{Sanders} \& {McGaugh}(2002)}]{2002ARA&A..40..263S}
{Sanders}, R.~H. \& {McGaugh}, S.~S. 2002, \araa, 40, 263

\bibitem[{{Seigar} {et~al.}(2008){Seigar}, {Barth}, \&
  {Bullock}}]{2008MNRAS.389.1911S}
{Seigar}, M.~S., {Barth}, A.~J., \& {Bullock}, J.~S. 2008, \mnras, 389, 1911

\bibitem[{{Sofue} \& {Rubin}(2001)}]{2001ARA&A..39..137S}
{Sofue}, Y. \& {Rubin}, V. 2001, \araa, 39, 137

\bibitem[{{Spergel} {et~al.}(2003){Spergel}, {Verde}, {Peiris}, {Komatsu},
  {Nolta}, {Bennett}, {Halpern}, {Hinshaw}, {Jarosik}, {Kogut}, {Limon},
  {Meyer}, {Page}, {Tucker}, {Weiland}, {Wollack}, \&
  {Wright}}]{2003ApJS..148..175S}
{Spergel}, D.~N., {Verde}, L., {Peiris}, H.~V., {et~al.} 2003, \apjs, 148, 175

\bibitem[{{Stark} \& {Binney}(1994)}]{1994ApJ...426L..31S}
{Stark}, A.~A. \& {Binney}, J. 1994, \apjl, 426, L31

\bibitem[{{Teuben}(1995)}]{1995ASPC...77..398T}
{Teuben}, P. 1995, in Astronomical Society of the Pacific Conference Series,
  Vol.~77, Astronomical Data Analysis Software and Systems IV, ed. R.~A.
  {Shaw}, H.~E. {Payne}, \& J.~J.~E. {Hayes}, 398--+

\bibitem[{{Unwin}(1983)}]{1983MNRAS.205..773U}
{Unwin}, S.~C. 1983, \mnras, 205, 773

\bibitem[{{van den Bosch} {et~al.}(2000){van den Bosch}, {Robertson},
  {Dalcanton}, \& {de Blok}}]{2000AJ....119.1579V}
{van den Bosch}, F.~C., {Robertson}, B.~E., {Dalcanton}, J.~J., \& {de Blok},
  W.~J.~G. 2000, \aj, 119, 1579

\bibitem[{{van der Marel} \& {Cioni}(2001)}]{2001AJ....122.1807V}
{van der Marel}, R.~P. \& {Cioni}, M.-R.~L. 2001, \aj, 122, 1807

\bibitem[{{Verley} {et~al.}(2009){Verley}, {Corbelli}, {Giovanardi}, \&
  {Hunt}}]{2009A&A...493..453V}
{Verley}, S., {Corbelli}, E., {Giovanardi}, C., \& {Hunt}, L.~K. 2009, \aap,
  493, 453

\bibitem[{{Walterbos} \& {Braun}(1994)}]{1994ApJ...431..156W}
{Walterbos}, R.~A.~M. \& {Braun}, R. 1994, \apj, 431, 156

\bibitem[{{Walterbos} \& {Kennicutt}(1987)}]{1987A&AS...69..311W}
{Walterbos}, R.~A.~M. \& {Kennicutt}, Jr., R.~C. 1987, \aaps, 69, 311

\bibitem[{{Walterbos} \& {Kennicutt}(1988)}]{1988A&A...198...61W}
{Walterbos}, R.~A.~M. \& {Kennicutt}, Jr., R.~C. 1988, \aap, 198, 61

\bibitem[{{Widrow} {et~al.}(2003){Widrow}, {Perrett}, \&
  {Suyu}}]{2003ApJ...588..311W}
{Widrow}, L.~M., {Perrett}, K.~M., \& {Suyu}, S.~H. 2003, \apj, 588, 311

\bibitem[{{Worthey} {et~al.}(2005){Worthey}, {Espa{\~n}a}, {MacArthur}, \&
  {Courteau}}]{2005ApJ...631..820W}
{Worthey}, G., {Espa{\~n}a}, A., {MacArthur}, L.~A., \& {Courteau}, S. 2005,
  \apj, 631, 820

\end{thebibliography}
\end{document}